\documentstyle[12pt]{article}
\input{psfig}

\def\nn{\nonumber}

\newcommand{\be}{\begin{equation}}
\newcommand{\ee}{\end{equation}}
\newcommand{\ba}{\begin{eqnarray}}
\newcommand{\ea}{\end{eqnarray}}
\newcommand{\ve}{\varepsilon}
\newcommand{\nit}\noindent
\newcommand{\D}\partial

\newcommand{\lan}\langle
\newcommand{\ran}\rangle

\newcommand{\tA}{\tilde A}

\newcommand{\tB}{\tilde B}

\newcommand{\tC}{\tilde C}

\newcommand{\Z}{{\rm Z\kern-.35em Z}}
\newcommand{\bP}{{\rm I\kern-.15em P}}
\newcommand{\Q}{\kern.3em\rule{.07em}{.65em}\kern-.3em{\rm Q}}
\newcommand{\R}{{\rm I\kern-.15em R}}
\newcommand{\h}{{\rm I\kern-.15em H}}
\newcommand{\C}{\kern.3em\rule{.07em}{.55em}\kern-.3em{\rm C}}
\newcommand{\T}{{\rm T\kern-.35em T}}
\newtheorem{theo}{Theorem}[section]

\newtheorem{cor}{Corollary}[section]
\newtheorem{rmk}{Remark}[section]
\begin{document}

\title{\bf Dynamics of lattice kinks}
\author{
P.G. Kevrekidis \thanks{Department of Physics and Astronomy, 
 Rutgers University, 
 Piscataway, NJ 08854-8019 USA} \hspace{.05 in}
 and
  M.I. Weinstein \thanks{ Mathematical Sciences Research, Bell Laboratories -
 Lucent Technologies, and Department of Mathematics, University of Michigan - Ann Arbor, 600 Mountain Avenue, Murray Hill,
 NJ 07974-0636 USA
   }}
\date{\today} 
\maketitle

\begin{abstract}
 We consider a class of Hamiltonian 
 nonlinear wave equations governing a field defined
on a spatially discrete one dimensional lattice, with discreteness 
parameter, $d=h^{-1}$, where $h>0$ is the lattice spacing.
The specific cases we consider in detail are the discrete sine-Gordon (SG)
and discrete $\phi^4$ models.  
  For finite $d$ and in the continuum limit ($d\to\infty$) 
 these equations have static kink-like (heteroclinic) states which are
stable. In contrast to the continuum case, due to the breaking of 
Lorentz invariance, discrete kinks cannot be ``Lorentz boosted" 
 to obtain traveling discrete kinks. Peyrard and Kruskal pioneered the study
of how a kink, initially propagating in the lattice dynamically 
 adjusts in the absence 
of an available family of 
 traveling kinks.
We study in detail the final stages of the discrete kink's evolution during
 which it is pinned to a specified lattice site (equilibrium position in the 
Peierls-Nabarro barrier). 
  We find:

\noindent (i) for $d$ sufficiently large
 (sufficiently small lattice spacing),   
 the state of the system approaches
 an asymptotically stable ground state  static kink
(centered between lattice sites).

\noindent (ii) for $d$ sufficiently small $d<d_*$ the static 
 kink bifurcates to one or more  
 time periodic states. 
For the  discrete $\phi^4$ we have: wobbling
kinks which have  the same  spatial symmetry 
as the static kink as well as  ``g-wobblers'' and ``e-wobblers'', 
which have different spatial symmetry. In the discrete sine-Gordon
case, the ``e-wobbler'' has the spatial symmetry of the kink 
whereas the ``g-wobbler'' has the opposite one.
These time-periodic states may be regarded as a class of discrete
breather / topological defect states; they are spatially localized and
time periodic oscillations mounted on a static kink background.

The large time limit of solutions with initial data near a kink 
is marked by  damped oscillation about
one of these two types of asymptotic states.
In case (i) we compute the
characteristics of the damped oscillation (frequency and $d$- dependent 
 rate of decay).
In case (ii) we prove the existence of, and give
analytical and numerical evidence for the asymptotic 
 stability of wobbling solutions.

The mechanism for decay is the radiation  of excess energy,
 stored in {\it internal modes},  
 away from the kink core to infinity. 
This process is studied in detail using general techniques of 
 scattering theory and 
 normal forms.
 In particular, we derive a  
{\it dispersive normal form}, from which one can anticipate the 
character of the dynamics.
 The methods we use are very general and are
  appropriate for the study of dynamical
systems which may be viewed as a system of discrete oscillators 
 ({\it e.g.} kink together
with its {\it internal modes}) coupled to a field ({\it e.g.} 
 dispersive radiation or 
 phonons). The approach is
  based on and extends an approach of one of the authors (MIW)
and A. Soffer in previous work.  
 Changes in the character of the dynamics, as $d$ varies, are manifested in 
topological changes in the phase portrait of 
the normal form.
 These changes are due to changes in the types of resonances 
which occur among the discrete internal modes and the continuum
 radiation modes, as $d$ varies.

Though derived from a time-reversible dynamical system, this
 normal form has a dissipative character. The dissipation is of an internal 
nature, and corresponds to the transfer of 
energy from the discrete to continuum radiation modes.  The coefficients 
which characterize the time scale of damping (or {\it lifetime} of the internal
mode oscillations)  
 are a nonlinear analogue of ``Fermi's golden rule", which arises in the theory 
of spontaneous emission in quantum physics. 

\end{abstract}


\section{Introduction}
\setcounter{equation}{0}

Coherent structures, {\it e.g. kinks, solitary waves, vortices},
 play a central role, as carriers of energy,
 in many  physical systems. An understanding of their dynamical properties,
 {\it e.g.} stability, instability, metastability, is an important
problem. 
While for many years  Hamiltonian partial differential equations 
and their coherent structures, defined on a 
spatial continuum,    
 have received a great deal of attention, there
has been increasing interest in spatially discrete systems. Two important 
reasons are that (a) certain phenomenoma are intrinsically associated with 
discreteness and (b) numerical approximations of continuum systems involve
the introduction of discreteness, which may lead to spurious numerical phenomena which require recognition. 
Some examples of the use of discrete systems in the modeling of 
  physical phenomena are:
 the problem of 
dislocations propagating on a lattice (for which the
Frenkel - Kontorova or discrete sine-Gordon model was originally
proposed) \cite{FK,AC,H}, arrays of coupled Josephson junctions, 
\cite{UDVPHO,HGGBB,XLY}, 
the problem of the local denaturation of the DNA double helix
\cite{YZW,D,DPW,DPB} and coupled optical waveguide arrays \cite{dnlsexperiment,MPAES}. 

A result of the many investigations of discrete systems
   over the last ten to fifteen  years 
 has been a recognition, mainly through numerical
  experiments and heuristic 
arguments, of the often sharp contrast in behavior between 
 the dynamics of discrete
systems and their continuum analogues.
Some of these contrasts 
are easy to anticipate. For example, continuum systems modeling phenomena
in a  homogeneous
environment are translation invariant in space, and may have further symmetry,
{\it e.g.} Galilean or Lorentz invariance that enable one to construct 
traveling solutions from static solutions.  The analogous discrete system is 
expected to lose these symmetries and therefore the existence of traveling 
wave solutions is now brought into question.
A natural question concerns the propagation of energy in the lattice;
{\it 
 if one 
 initializes the system with  data which, 
 for the continuum model would result
in a coherent structure propagating through the continuum,  
  what is the corresponding behavior for the energy distribution on intermediate
and long times scales on the lattice? Does the system
  ``find" a coherent structure 
 to carry the energy?  Does the energy get trapped or pinned?} 
In this paper we seek to obtain insight into these questions for a class of
discrete nonlinear wave equations. The special cases we consider in detail 
are the discrete sine-Gordon (SG) and discrete $\phi^4$ equations.
The methods however are rather general and apply to systems which can be 
viewed as the interaction between a finite dimensional system of 
``oscillators'' with an infinite-dimensional system governing
a continuous spectrum of waves; see the further discussion below
and in section 7.

The basic characteristics of the dynamics of coherent structures in 
 discrete systems  
were systematically explored  
   in a pioneering paper by Peyrard and Kruskal \cite{PK}; see also the
contemporaneous papers \cite{IM,PA,WES}. 
 Of particular relevance to our work are the more recent articles
 of Boesch, Willis and coworkers
\cite{WES,SWEB,BSW,BW1,BW2,BWE}.
In this work the discrete sine-Gordon equation is solved with 
 kink-like initial data on the lattice.
  Observed is a rapid initial velocity adjustment of the kink, 
 a quasi-steady state phase, 
resonances of the kink oscillations with phonons (continuous spectral modes)
and the emission of radiation,
  which results in the kink's deceleration and   
 eventual pinning in the Peierls-Nabarro (PN) potential. 
The final stage involves the relaxation to an asymptotic state, which is  
either a static or time-periodically ``dressed" kink.
In a rough sense, this is a result of the absence of a smooth family 
of traveling kink solutions 
due to broken Lorentz invariance. Insight into this process can be 
gleaned by studying the linearized  spectrum about the kink.

An important feature of the discrete systems that makes them  
very different from their continuum analogues is the presence of 
additional neutral oscillatory modes in the spectrum  of the linearization
about the coherent structures. The sources of these {\it internal modes} 
are principally of two types; see for example \cite{BKP,KPCP,K,KJ}.
 (1) The continuum system is translation 
invariant, a symmetry which leads to the continuum linearization having 
zero modes, eigenmodes and generalized eigenmodes corresponding to zero 
frequency. Viewed as a perturbation of the continuum system, the discrete,
but ``nearly" continuum problem is expected to have nearby 
 modes to which these zero
modes have been perturbed. If the coherent state is stable, these modes should be
neutrally stable, {\it i.e.} that is, they must correspond to purely imaginary 
eigenvalues. (2) It is possible that discrete neutral modes may emerge from the 
continuous (phonon) spectrum.

Although some of these phenomena have been
identified in the  early numerical investigations  
\cite{CTBK}, there has not been a systematic dynamical systems study relating  
particular resonances to  
 the rate with which the kink is trapped 
  by a ``valley" in the PN potential or, once inside 
  such a valley,
the rate with which the kink relaxes to its asymptotic   
state. The main results in this
direction date from the work of Ishimori and Munakata
\cite{IM}, in which the McLaughlin-Scott direct perturbation 
scheme \cite{MS} is used,  
  valid in only in the nearly continuum regime,  
 and the work of Boesch, Willis and El-Batanouny
\cite{BSW},  which is  based on numerical simulations and heuristic arguments.

In this paper, we introduce a systematic approach to the study  
of these phenomena. The work is based on 
the recent
work of one of us (MIW) with A. Soffer on a time dependent theory of
metastable states in the context of: 
quantum resonances \cite{GAFA},
ionization type problems (parametrically excited Hamiltonians) 
\cite{SWjsp} and 
resonance and radiation
 damping of bound states in nonlinear wave equations \cite{SWinvent}; see also
 \cite{KirrWeinstein, MSW}
A fruitful point of view adopted in these works and in the present 
 work is that the  dynamics can be understood as 
 the interaction between a finite
dimensional dynamical system, 
 governing the bound state (internal modes plus kink) 
part of the solution and an infinite dimensional dynamical system  
 and governing 
radiative behavior. 
Using the tools of scattering theory and the idea of normal forms, we derive
a {\it dispersive normal form}, which is a closed (up to controllable
error terms) finite dimensional system governing the internal mode components
of the perturbation. From this normal form many aspects of the  
the large time asymptotic state are deducible.

 The discrete systems we study
  ({\it e.g.} discrete sine-Gordon and discrete $\phi^4$) depend on a 
discreteness parameter, $d$; as $d$  increases the
  continuum limit is approached.
 The character of the normal form (topological character of its phase
 portrait) changes with $d$ because the  
 kinds of resonances 
which occur among discrete mode oscillations and continuum radiation 
 (classifiable in terms of integer linear combinations of internal mode 
frequencies) change with $d$. 


 Our analytical and numerical studies lead us to the following 
picture concerning the 
  dynamics in a neighborhood of the kink: 
 
{\it 
\noindent (1) The ground state kink, $K_{gs}$, is always Lyapunov stable
\footnote{ If the initial data is in a small neighborhood of $K_{gs}$, 
 then the solution 
remains in a small neighborhood of $K_{gs}$ for all time.
  This notion of stability does not
 however imply convergence to 
$K_{gs}$.}

\noindent (2) If $d$, the discretization parameter,
  is sufficiently large (approaching the spatial continuum 
case), the ground state kink 
is an attractor, {\it i.e.} is asymptotically stable.
\footnote{
By asymptotic stability we mean that a small perturbation of the kink
gives rise to a
  solution which converges as $t\to\pm\infty$, in some physically relevant norm,
 to a kink. Asymptotic stability is a notion of stability commonly
associated with dissipative dynamical systems and is
 not commonly associated with general
energy conserving systems. In this work we are studying infinite dimensional
 Hamiltonian
systems on infinite spatial domains. These systems have the possibility of
 radiation of energy to infinity,
 while keeping the total energy of the system preserved.
 Thus dissipative behavior
is realized through dispersion and eventual radiation of energy
 out of any compact
 set; see, for example, the short overview in \cite{WContemp} and references
therein.}
 In this case the kink is approached at different algebraic rates
 depending on the range of $d$ values. We infer this from 
the normal form analysis of section 4.

\noindent (3) For $d$ sufficiently small, our   
 discrete nonlinear wave models have one or more branches of 
 finite energy time-periodic solutions which 
bifurcate from $K_{gs}$; see Theorem 5.1 and Corollary 5.1. 
 For the specific cases of discrete SG and discrete $\phi^4$ our results 
imply, for various regimes of $d$, that there exist
 {\it wobbling kinks}, $W$. 
 These are time periodic solutions with the same spatial symmetry as the kink
and have been previously observed in numerical simulations. Additionally,
our results imply in certain regimes of $d$, the existence of time periodic  
 solutions (such as $gW$, $eW$ in the $\phi^4$ and $gW$ in the SG), $u(t)$, with the property that 
 $u(t)-K_{gs}$ is, to leading order
in the direction of an {\it even} internal mode. Tables 3 and 7 in section
 4 indicate the regimes in which these various solutions denoted
 occur. The wobbling solutions ($W,gW,eW$) may be 
viewed as a class of discrete
breather / topological defect states. They are spatially localized and
time-periodic oscillations mounted on a static kink background. The usual
discrete breathers \cite{AUB,MA} are mounted on a zero background.
This is analogous to the situation with solitons. Dark solitons of the
defocusing nonlinear Schr\"odinger (NLS) equation are mounted on a
non-zero continuous wave background, while standard solitons of focusing 
NLS sit on a zero background.

\noindent (4) Numerical simulations and the normal form analysis 
 indicate that in various $d$ regimes
 these time periodic states are local attractors for the dynamics.

\noindent(5)  The analysis of section 5 implies some nonexistence 
results concerning solutions of discrete nonlinear wave equations 
in a neighborhood of the kink. We have that 
in certain regimes of the discreteness parameter, $d$, no 
nontrivial time-periodic solution exists, and that 
  time quasiperiodic solutions do not exist 
 in any regime of $d$. On the other hand, the normal form analysis and
  numerical simulations indicate, in some regimes
of $d$, that periodic or quasiperiodic oscillations can be very long lived;
see section 6, and in particular, the discussion of Regime VI in section 6.1.
Thus, we may think of the system as possessing {\it metastable}
periodic and quasi-periodic solutions. In regimes of $d$ where we show  
 that the wobbling
 kink, $W$, is unstable on long time scales due to a resonance of the 
``shape mode" with continuum radiation modes, we have the analogue of the 
 the wobbling solution of continuum $\phi^4$, proved to be stable 
on large but finite time scales by Segur \cite{Segur}. On an 
 infinite time scale this wobbling solution behaves as those of the 
discrete system for large enough $d$; eventually the oscillations damp 
at an algebraic rate leaving a kink in the limit \cite{Manton,PSW}. 
 For the continuum system,
the limiting kink may be traveling, while for the discrete system it is a static
ground state (centered between lattice sites) kink.   
}

In (2) the dispersive normal form has a dissipative character; the effect
of coupling the internal mode oscillations to radiation is modeled 
by an appropriate nonlinear friction; 
 see section 4 and the discussion in the introduction
to \cite{SWinvent} on another related model. The {\it damping} or 
 {\it friction coefficients} 
 are given 
by formulae which can be understood as a nonlinear generalization of {\it Fermi's
golden rule}, arising in the context of the theory of spontaneous emission in
atomic physics. 
Finally, it is worth noting that we obtain a dissipative and therefore apparently
 time-irreversible normal form from 
 a system which is conservative and time-reversible.
  There is no contradiction because the dissipation 
is of an internal nature; it signifies the transfer of energy from the discrete
internal mode oscillations to the continuum dispersive waves which propagate to 
infinity. That dissipative dynamical systems emerge from conservative systems 
 which are a coupling of a low
dimensional dynamics  to infinite dimensional dynamics (``masses and springs
  coupled to strings") has been observed in many contexts;
 see, for example, \cite{Lamb,BK,L} and the discussion in 
 \cite{GAFA,SWjsp,SWinvent}. 

The paper is  organized as follows: 
\begin{itemize}
\item Section 2 begins with a general discussion of discrete nonlinear 
 wave equations which support kink-like (heteroclinic) structures and then 
specializes to a discussion of the discrete sine-Gordon (SG) and discrete 
 $\phi^4$ systems. 
\item In section 3, we present the decomposition of solutions into 
 discrete internal mode components and radiative components, enabling 
us to view the dynamics near a kink as the interaction of finite and infinite
 dimensional Hamiltonian systems.
\item In section 4 the dispersive normal forms are derived and discussed
 for both discrete SG and discrete $\phi^4$.
\item In section 5, we prove 
that if $\Omega$ is an internal mode frequency and no multiple of it lies
in the continuous spectrum of the kink (phonon band), then there is a family 
of finite energy time periodic solutions which bifurcate from the kink
in the ``direction" of the corresponding internal mode. 
 The proof is based on 
 the {\it Poincar\'e continuation} method and is an application of the implicit
function theorem in an appropriate Banach space. This approach was used also
by MacKay and Aubry \cite{MA}, who constructed discrete breathers of nonlinear
 wave equations in the 
{\it anti-integrable} limit. 
\item In section 6 we combine the normal form analysis of section 
 4 and existence theory for periodic solutions of section 5 with observations 
based on numerical simulations to obtain a detailed picture of the dynamics
in a neighborhood of the ground state kink.
\item Finally, in section 7, we summarize our results
and give directions of interest for future research.
\end{itemize}
\medskip

\subsection{Notation}

\begin{itemize}
\item
 $\Z$ denotes the set of all integers and $\R$ denotes the set of real numbers.

\item 
 For $u=\{u_i\}_{i\in\Z}$, $\delta_h^2u$ denotes the discrete Laplacian of $u$
defined by:
 \be (\delta^2_hu)_i = h^{-2}\left( u_{i+1}-2u_i+u_{i-1}\right).
\nn\ee
In the case $h=1$ we shall use the simplified notation $\delta^2=\delta_1^2$.

\item The inner product of vectors $u$ and $v$ in $l^2(\Z)$ is given by
\be
\lan u,v \ran\ =\ \sum_j \bar u_jv_j.\nn\ee

\item The $t$ subscript denotes time partial derivatives, {\it e.g.}
\ $u_t(t) = \partial_t u(t)$, whereas $i$ denotes the lattice
site numbering.

\item $l^2(\Z)$ is the Hilbert space of sequences $\{u_i\}_{i\in\Z}$
 which are square summable.

\item $H^s(I)$ denotes the Hilbert space of functions $f$, 
 defined on $I\subset\R$
 such that $f$ and all its derivatives of order $\le s$ are square-integrable.
For the case of $2\pi$ periodic functions, $I=S^1_{2\pi}$, an equivalent 
 norm on $H^s$ is given by:
\be \| f\|^2_{H^s}= \sum_{n\in\Z}(1+|n|^2)^s |f_n|^2,\nn\ee
where $f_n$ denotes the $n^{th}$ Fourier coefficient of $f$:
\be f_n = (2\pi)^{-{1\over2}}\int_0^{2\pi}e^{-2\pi i nt}f(t)\ dt.\nn\ee

\item For example, $H^2(\R;l^2(\Z))$ denotes the space of functions
 $f(t,i)$ which are $H^2$, as functions of $t$, with values in the space of 
$l^2(\Z)$ functions.

\item $\sigma(B)$ and $\sigma_{cont}(B)$ denote respectively the spectrum and 
 the continuous ({\it phonon}) spectrum of the
operator $B$.

\item $0<d$ denotes the {\it discretization parameter}; 
 $d$ large is the  spatial 
continuum regime, while for $d$ small the effects of discreteness are 
strong.

\end{itemize}

\section{ Discrete and continuum wave equations}
\setcounter{equation}{0}

\subsection{General Background}

Two model nonlinear wave equations on lattices
 that have played a central role in the 
theory of nonlinear waves are the {\it discrete sine-Gordon equation} (SG) 
 and the 
{\it discrete $\phi^4$ model} ($\phi^4$).  
 These equations may be viewed as governing the 
dynamics of a  chain of unit mass particles which, in equilibrium,
  are equally spaced, a unit
distance apart. The particles are then subjected to a conservative force derived 
from a potential. In terms of the displacement $u_i$ of the $i^{th}$ particle 
from equilibrium the equations of motion are\footnote{We use the notational
 conventions introduced by Peyrard and Kruskal \cite{PK}.}: 
\be
u_{i,tt}\ =\ (\delta^2u)_{i}\ -\ d^{-2} V'(u_i).
\label{eq:discretemodels}
\ee
The case of SG and $\phi^4$ correspond, respectively, the choices of potential: 
\ba
V(u) &=& 1-\cos u,\ \ ({\rm SG})\nn\\
V(u) &=& {1\over4}(1-u^2)^2.\ \ (\phi^4)\nn
\ea
More generally, we assume $V$ has three continuous derivatives and satisfies 
certain constraints appearing below, related to the existence of kink solutions 
 or homoclinic  orbits.
     
The parameter $d$ is a fixed constant with, $d^{-2}$, having
the interpretation of 
the ratio of on- site potential energy to elastic
coupling energy. 
 
\noindent{\bf Relation between discrete and continuum systems:}
 The parameter $d$ can also be seen to play the role of the reciprocal
of the lattice spacing in  passing between continuum and  discrete models. 
Consider 
 the continuum model governing the displacement $v(T,x)$:
\be v_{TT}= v_{xx} - V'(v).\label{continuummodels}\ee
Introducing the lattice spacing parameter, $h$, and replacing 
$u_{xx}(x)$ by $(\delta^2u)_i$ we obtain the discrete nonlinear wave equation
 governing $v_i = v(T,i;h)$:
\be v_{i,TT}= (\delta_h^2v)_i - V'(v_i).\nn\ee
The form of the discrete nonlinear wave equation we use is obtained 
by setting $t=h^{-1}T$ and defining the time-scaled displacement $u(t,i;h)=v(T,i;h)$.
Then, $u(t,i;h)$ satisfies (\ref{eq:discretemodels}) with $d=h^{-1}$. 

Taking the formal limit $d\uparrow\infty$ or equivalently
 $h\downarrow0$ we have that 
\be U(T,i)\equiv\lim_{d\to\infty} u(dT,i;d^{-1})=
\lim_{d\to\infty} v(T,i;d^{-1})\nn\ee
satisfies the 
 associated {\it continuum} nonlinear wave equation (\ref{continuummodels}). 

\noindent{\bf Hamiltonian structure:}
  The discrete and continuum
 equations we consider are infinite dimensional Hamiltonian systems
with  Hamiltonian energy functionals\footnote{ There is freedom 
 in the choice of potential $V$;
 we choose $V$ so that the Hamiltonian energy of
the static kink is finite.}:
\be
{\cal H} = \sum_i {1\over2}u_{i,t}^2+{1\over2}\left(u_{i+1}-u_i\right)^2
 + d^{-2} V(u_i)\label{eq:denergy}\ee
in the discrete case and
\be
{\cal H} = \int_\R \frac{1}{2}(u_T^2+ u_x^2) + V(u)\ dx.\label{eq:energy}
\ee

\noindent{\bf Static kink solutions:}
 Consider time-independent or {\it static} solutions of these 
dynamical systems. In each case, such solutions satisfy the equation obtained
by setting time derivatives equal to zero. Thus, in the discrete case we have:
\be (\delta^2 K)_i = V'(K_i),\label{eq:dstatic}\ee
and in the continuum case
\be K''(x) = V'(K(x)) \label{eq:static}\ee
Spatially uniform solutions occur at the critical points of the potential, $V$.
Of particular interest are the static {\it kink} solutions. These are static 
solutions which are {\it heteroclinic}. That is, they are static 
 solutions which can be viewed as connections, as $i\to\pm\infty$ 
 (respectively $x\to\pm\infty$)  in the phase space of 
 two distinct "unstable" 
 equilibria, values $K^\pm_*$  for which 
\ba V(K_*^+)&=&V(K_*^-),\  
 V'(K^\pm_*)=0\nn\\
  V_*''&\equiv& V''(K^+_*)=V''(K^-_*)>0.\label{eq:potential}\ea

\begin{figure}[ptbh]
\centerline{\psfig{width=4in,height=3.9in,file=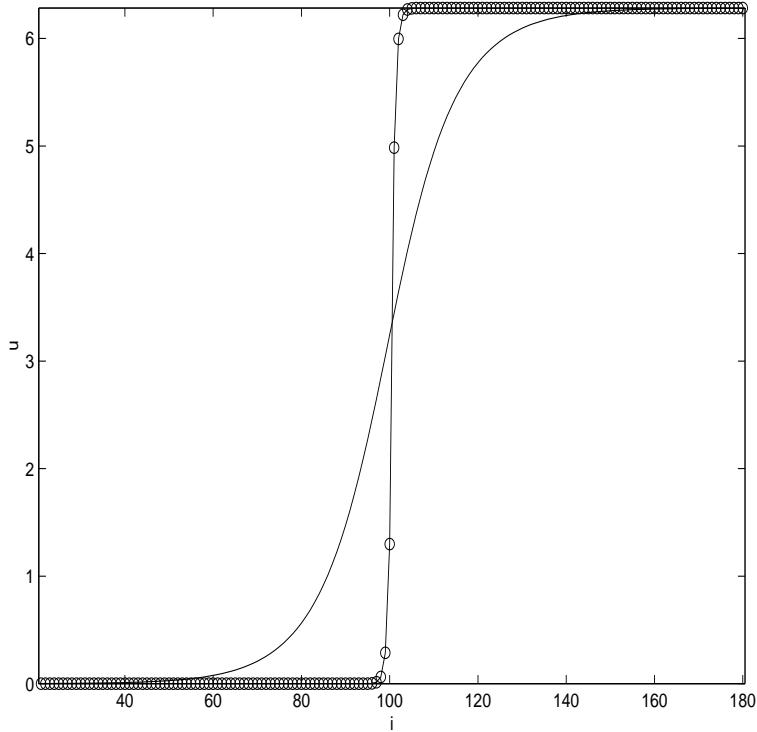}}
\caption{Ground state static kink for the very strongly discrete
 sine-Gordon model (--o--) $d=0.6$, and very close to the continuum
limit $d=10$.}
\end{figure} 

Kink solutions can also be constructed by variational methods. Consider the 
Hamiltonian energy functional, ${\cal H}$ restricted to $t$-independent  
 functions:
\ba h[u]\ &=&\ \sum_i\left(\ {1\over2}(u_{i+1}-u_i)^2 + d^{-2}V(u_i)\ \right),\
 {\rm discrete\ case}\label{eq:staticham-d}\\
    h[u]\ &=&\ \int_\R\ \frac{1}{2} u_x^2+ V(u)\ dx\ 
 {\rm continuum\ case.}\label{eq:staticham-c}\ea  
Then, the Euler Lagrange equations associated with $h[u]$ is the equation 
 for the kink, (\ref{eq:dstatic}). That is, 
 if $K_{gs}$ denotes a minimizer of $h[u]$ then since
 for all $\psi\in l^2(\Z)$,
\be {d\over d\tau}\ h[K_{gs}+\tau\psi]\left|_{\tau=0}\right.=0.\nn\ee
This implies that for all $\psi\in l^2(\Z)$
\be
\sum_i \ \left[-(\delta^2K_{gs})_i + d^{-2}V'(K_{gs,i})\right]\psi_i\ =\  0.
\nn\ee
This is equivalent to  (\ref{eq:dstatic}).
A solution constructed by minimization of $h[u]$ is called a 
 {\it ground state kink}.\footnote{ A proof that the minimum 
  of the functional (\ref{eq:staticham-d}) 
is attained can be given using the following strategy, used in a similar problem
\cite{rockymountain}:
  For positive integers, $N$, define
 $h_N[u]$ to be the truncated Hamiltonian,
 where the summation is taken over $-N\le n\le N$.
 Consider $h_N$, restricted to vectors satisfying $u_{\pm N}=K_*^\pm$.
 For any admissible vector, $u=\{u_i\}_{|i|\le N}$, it is 
 possible, by rearrangement, to replace it with another, $v$, which is
monotonically increasing and for which $h[v]\le h[u]$. 
Therefore the minimizer of $h_N[\cdot]$, $K_{gs}^{(N)}$,  is 
 monotonically increasing from $K_*^-$ to $K_*^+$.  
 \'A priori estimates, derived 
from the Euler-Lagrange equation (\ref{eq:dstatic}),  
and the boundary conditions enable one to pass to the limit and 
 obtain a minimizer of $h[\cdot]$, $K_{gs}=\{K_{gs,i}\}_{i\in\Z}$,
 a monotonically 
increasing static kink.
}

\noindent{\bf Invariance and broken invariance:} 

An important structural 
 difference between the discrete and continuum cases is that the 
continuum case has greater symmetry. In particular, (\ref{continuummodels})
has the property of translation and 
   Lorentz invariance, while (\ref{eq:discretemodels}) 
does not. A consequence of this is that static solutions
  of (\ref{continuummodels}) can be translated:
\be K(x)\mapsto K(x-x_0)\nn\ee
to give a recentered kink and 
Lorentz boosted:
\be K(x)\mapsto K((x-vt)/\sqrt{1-v^2}),\ |v|<1,\nn\ee
 to give a traveling solution of velocity $v$.
 These symmetries are absent in the discrete models. Discrete models do however have a discrete translation symmetry but in the models considered this does not
give rise to discrete traveling waves.

\noindent{\bf Dynamic stability of kinks:}
 The issue of dynamic stability is a subtle one. 
First, there is the question 
of what is the object we expect to be stable. 
Also, since
  our systems are infinite dimensional and not all norms are equivalent, 
 different choices of norms 
will measure different phenomena. 
Here we briefly discuss three related notions of stability we have in mind: (a)
orbital  Lyapunov stability, (b) asymptotic stability and (c) linear spectral
 stability.

\noindent (a) {\it Orbital Lyapunov 
 stability}  means stability of the shape of the kink; if at $t=0$ the data
is nearly shaped like a kink then it remains shaped like a kink for all $t\ne0$.
The kink-like part of the solution will typically move so the solution at 
different times is near some time-dependent spatial translate of the static 
 kink (element of the symmetry group orbit of the static kink). 

In the continuum case, orbital Lyapunov stability in the space $H^1(\R)$
 is a consequence of characterization of the kink as a local minimizer of ${\cal H}$. A detailed proof is presented in \cite{HPW}. A simple proof of Lyapunov
stability of ground state discrete kinks can be given based on the same ideas.
\footnote{The proof is based on the following idea. Since $K_{gs}$ is an energy 
minimizer, by (\ref{eq:Bge0}), the second variation at the kink, $B^2$ 
 is non-negative. In fact, it is strictly positive; that is, 
 for all $\psi\in l^2(\Z)$, 
 $\lan\psi,B^2\psi\ran \ge \omega_g^2\lan\psi,\psi\ran$,
 where $\omega_g^2=\omega_g^2(d)>0$ is the discrete eigenvalue to which the 
 zero mode of the  continuum equation perturbs due to discretization of space.
 Suppose we have initial data near a kink: $u(0) = K_{gs} + v_0,\ \D_tu(0)=v_1$
 and let $u(t)=K_{gs} + v(t)$ be the resulting solution. Assume that 
$\|v_0,v_1\|_{l^2}\sim\ve$, where $\ve$ is sufficiently small. 
 Then, since $K_{gs}$ is a critical point of $h[\cdot]$ we have:  
\ba \ve\sim h[K_{gs}+v(t)]|_{t=0}-h[K_{gs}] &=& h[K_{gs}+v(t)]-h[K_{gs}] 
  = {1\over2} \lan v(t),B^2 v(t)\ran + {\cal O}(\|v(t)\|_{l^2}^3)\nn\\
  \ge {\omega_g^2\over2} \|v(t)\|_{l^2}^2 -c\|v(t)\|_{l^2}^3.\nn\ea
implying that $\|v(t)\|_{l^2}\sim {\cal O}(\ve)$ for all $t\ne0$, and
 $K_{gs}$ is stable.}

\noindent (b) {\it Orbital asymptotic stability}
 means that if the initial data is nearly a kink
then  the solution converges to kink (possibly traveling, in the continuum case
  or pinned in the discrete case) as $t\to\infty$.

For both notions (a) and (b) we see that the stable object is the {\it family}
of solitary wave solutions. That is, in the case where there is a 
 multiparameter family of solutions (kinks, solitary waves ...), 
 it is the collection of all such
that is the stable object. Concerning the terminology {\it orbital stability},
 this family of solutions is related to the set of functions 
 generated by applying elements of the equation's symmetry group to the 
 solution (here, static kink), and thus we can think of the group orbit 
 as stable.

\noindent (c) {\it Spectral stability:}  

As is well known, dynamic stability of the particular solution 
  is related to the spectrum of the 
 linearization of the dynamical system about this solution.
Linearization about the kink gives an evolution equation 
 for infinitesimal perturbations of the following  form:
\be \left(\D_t^2+B^2\right)p = 0.\nn\ee
For the discrete case:
\be B^2p_i\ =\ -(\delta^2p)_{i}\ +\ d^{-2} V''(K_i)p_i,\nn\ee
and for the continuum case:
\be B^2p\ =\ -p_{xx} +  V''(K(x))p.\nn\ee

 At a minimum we expect that stability can hold only if no solutions 
of the linearized evolution equation, 
  corresponding to finite energy initial conditions,
can grow as $|t|$ increases. 
In carrying out linear stability analysis we seek solutions of the form
$p_i=e^{\lambda t} P_i$ (respectively, $p=e^{\lambda t}P(x)$)
and obtain linear eigenvalue problems of the form:
\be \left(B^2+\lambda^2\right)P\ =\ 0.\label{evp}\ee
Explicitly, we have:
\ba
-\lambda^2P_i\ &=&\ -(\delta^2P)_{i}\ +\ d^{-2} V''(K_i)P_i,\ 
{\rm discrete\ case}\nn\\
 -\lambda^2P\ &=&\ -P_{xx} +  V''(K(x))P.\ 
\  {\rm continuum\ case}\nn\ea

We say {\it $\lambda$ is in the $l^2$ spectrum (respectively, $L^2$ spectrum)
of the linearization about the kink}, or simply {\it spectrum of the kink} if   
 the operator $B^2+\lambda^2$ does not have a bounded inverse on $l^2$ 
 (respectively $L^2$). The spectrum will typically consist of two parts:
 (i) the point spectrum consisting of isolated 
 eigenvalues of finite multiplicity,
 for which the corresponding solution of (\ref{evp}) is in the Hilbert space,
and due to the infinite extent of the spatial domain 
 (ii) continuous spectrum, whose corresponding solutions ({\it 
 radiation or phonon modes})
 of (\ref{evp})
are bounded and oscillatory over the entire spatial domain. 

Note that the eigenvalue equation (\ref{evp}) has the following symmtery:
 if  $\lambda$ has the property that  (\ref{evp}) has a nontrivial solution
 solution 
 then  $-\lambda, \bar\lambda$ and $-\bar\lambda$ also have this property.
To avoid exponential growing solutions, we must require
that the linearized spectrum is a subset of the imaginary axis.
If this holds we say the solution is {\it spectrally stable}.

\noindent{\bf Spectrum of the kink:}

We are interested in the set of $\lambda$'s for which the eigenvalue equation
\be (B^2+\lambda^2)P=0\nn\ee
has a nontrivial solution. We begin by discussing the spectrum of $B^2$
and then take square roots to a obtain the spectrum of the the linearization 
about the kink.

 We now make two general remarks about the spectrum of the  
 kink solution, one concerning the continuous spectrum 
 and one concerning
the point spectrum.

\noindent {\it Continuous spectrum:}
The continuous
spectrum is determined by the solutions of the constant coefficient equation
obtained from (\ref{evp}) by evaluating its coefficients at spatial infinity.
In the discrete case we obtain:
\be -\lambda^2P_i\ =\ -(\delta^2P)_{i}\ +\ d^{-2} V_*''P_i,\nn\ee
where  $V_*''=\lim_{i\to\pm\infty} V''(K_i)$. This constant coefficient
equation can be solved in terms of exponentials which yield the  
character of the exact continuum eigensolutions of (\ref{evp}).
We let $\lambda = i\omega$. We find bounded oscillatory solutions of the form:
 $P_n=\exp(\sqrt{-1}kn),\ n\in\Z$, $k$ real and arbitrary,
  where satisfies the dispersion relation:
\be \omega^2 = 4\sin^2(k/2) + d^{-2}V_*''.\label{dispersionrelation}\ee
It follows that the continuous spectrum of $B^2$ is the positive interval
from $d^{-2}V_*''$ to $d^{-2}V_*''+4$.
Therefore, the continuous spectrum of the linearization about the kink solution
 consists of two intervals on the imaginary
axis: $ \pm i[d^{-1}\sqrt{V_*''},\sqrt{4+d^{-2}V_*''}]$.

\noindent{\it Point spectrum:} 
An important conclusion, concerning the point 
 spectrum of kink for the 
continuum equation, can be made as a consequence of the equation's 
 symmetry group. 
 If $K(x)$ is a kink then translation invariance implies that $K'(x)$ is a zero mode, a solution of the linear eigenvalue problem with 
eigenvalue zero, {\it $B^2 K=0$}.
  This zero mode is often called the {\it Goldstone mode}. Since the 
 eigenfunction, $K'$ does not change sign, zero is  
 the ground state  energy (lowest point in the point spectrum) 
 of the Sturm-Liouville operator
$-\D_x^2 + V''(K(x))$. This eigenvalue is of multiplicity one.

 If one views the discrete model as 
 a perturbation of the continuum  model, due to the absence of corresponding
symmetries in the discrete problem one expects the eigenvalue at zero 
 to move, as $d$ decreases from infinity ($h$, the lattice spacing, increases 
from zero), to the right of zero  or to the left  
 of zero. 
 In terms of the spectrum of the kink (plus or minus
 the square root of 
 the spectrum of $B^2$),  this means that 
  in the discrete case the spectrum of the kink consists of either (a) a
  purely imaginary pair, $\pm i\omega_s$
 (stable case) or (b) 
 a pair of real eigenvalues, symmetrically situated about the
origin (unstable case). We shall see both cases in 
our study of specific discrete models and we shall refer
to these eigenvalues and corresponding modes loosely as  Goldstone modes 
 as well. 

As we shall see in the specific models considered,
  other eigenvalues may appear in the gap between
the upper and lower branches of the continuous spectrum. Such modes of the 
linearization which lie in this gap are called {\it internal modes}, and their
number and location can change with $d$.
For SG we shall find that there are one or two pairs of internal modes,
 while for the $\phi^4$ model there are
two or three pairs of internal modes.

Finally, we remark that in the case of a {\it ground state kink}, one obtained 
by minimization of $h[u]$, the spectrum of $B^2$ is non-negative and the kink
is spectrally stable. To see this, let 
  $K_{gs}$ denote a minimizer of $h[u]$, which as 
 indicated above, is a critical point of $h[\cdot]$. Furthermore,    
 for all $\psi\in l^2(\Z)$,
\be {d^2\over d\tau^2}\ h[K_{gs}+\tau\psi]\left|_{\tau=0}\right.\ge0.\nn\ee
A calculation yields that this is equivalent to:
\be {1\over2}
 \lan\psi,B^2\psi\ran\ \ge\ 0,\ {\rm for\ all}\ \psi\in l^2(\Z).
 \label{eq:Bge0}\ee
 Therefore, $d^2h[K_{gs}]={1\over2}B^2$, the second variation of $h[\cdot]$ at the 
ground state kink, is a nonnegative self-adjoint operator. 
  It follows that the 
spectrum of the kink, $\pm i\sigma(B)$, lies on the imaginary axis.

\subsection{The Sine-Gordon equation}

Consider the discrete SG equation: 
\begin{eqnarray}
u_{i,tt}=(u_{i+1}+u_{i-1}-2u_i)-\frac{1}{d^2} \sin{u_i}.
\label{eq1}
\end{eqnarray}
A static  discrete $2\pi$ kink is a time-independent 
 solution, $\{K_i\}_{i\in\Z}$ 
  of (\ref{eq1}) which 
  satisfies the boundary conditions at infinity:
\be K_i\to 0,\ {\rm as}\  i\to -\infty,\ {\rm and}\ K_i\to 2\pi,\ {\rm as}\ 
  i\to\infty\nn\ee
 It is
well-known that there  
are two static kink solutions, a high energy  
one centered on a lattice
site and a low energy
 one centered between two consecutive lattice
sites \cite{PK,IM,CY,WES,SWEB,BSW,BW1,BW2,BKP,KPCP,K,KJ}. 
  The low energy kink corresponds to the 
minimizer of the  static Hamiltonian energy, $h[u]$, displayed in
(\ref{eq:staticham-d}). 
  As $d$ increases (the continuum limit)
the energy difference between the two static kink solutions, 
the so-called Peierls-Nabarro (PN) barrier, 
tends to zero and the scaled limit $u(i;d^{-1})=K_i$, with $i/d\equiv x$ 
fixed, converges to 
the continuum SG static kink solution 
\be K_{SG}(x)\ =\ 4 {\rm tan}^{-1}(e^x).\nn\ee
Linear stability analysis about the high energy and low
energy kinks was carried out in \cite{BCK}. 
From the general discussion of section 2.1, 
 the dispersion relation defining the continuous (phonon)  spectrum  
is  $\omega^2=d^{-2}+ 4\sin^2(k/2)$. Therefore, $B^2$ has a band of
 continuous
spectrum of length $4$,  $[d^{-2}, 4 + d^{-2}]$, and the 
 continuous  spectrum of the kink consists of the two bands,  
    $\pm i[d^{-1},\sqrt{4 + d^{-2}}]$  .

For the case of 
 the high energy kink, the point spectrum of $B^2$ contains a negative
eigenvalue derived from the zero (Goldstone) mode associated with the continuum
 (translation invariant) case. Since the eigenvalue parameter in (\ref{evp}) is
$-\lambda^2$, it follows that the spectrum of the  
high energy kink (parametrized by $\lambda$) consists of 
two discrete real (Goldstone) 
 eigenvalues, one positive and one negative, $\pm\omega_{us}$. These occur
  at  a distance of order
$exp(-\pi^2d)$ \cite{IM,PK,BTK,JKK}.

The low energy kink, minimizes the Hamiltonian. The second variation 
of $h[u]$ at the kink, $B^2$, is therefore a self-adjoint and 
  nonnegative operator. Its spectrum is therefore nonnegative and 
so the spectrum of the kink 
is purely imaginary. In this case, the Goldstone mode associated 
 with translation  invariance of the continuum problem gives rise to 
point spectrum consisting of a complex conjugate pair of eigenvalues 
with order of  magnitude  $\exp(-\pi^2d)$.
 Furthermore, for $d>d_e$, $d_e\sim 0.515$ a spatially antisymmetric {\it edge mode} 
of energy $\omega_e^2$  ($B^2e=\omega_e^2e$) 
 appears whose corresponding energy has a distance
  from the phonon band edge of 
order  ${\cal O}(d^{-4})$, for $d$ large \cite{KPCP,KJ}.
 This mode is not present in the continuum SG.

\begin{figure}[ptbh]
\centerline{\bf Spectrum of discrete SG ground state kink}
\centerline{\psfig{file=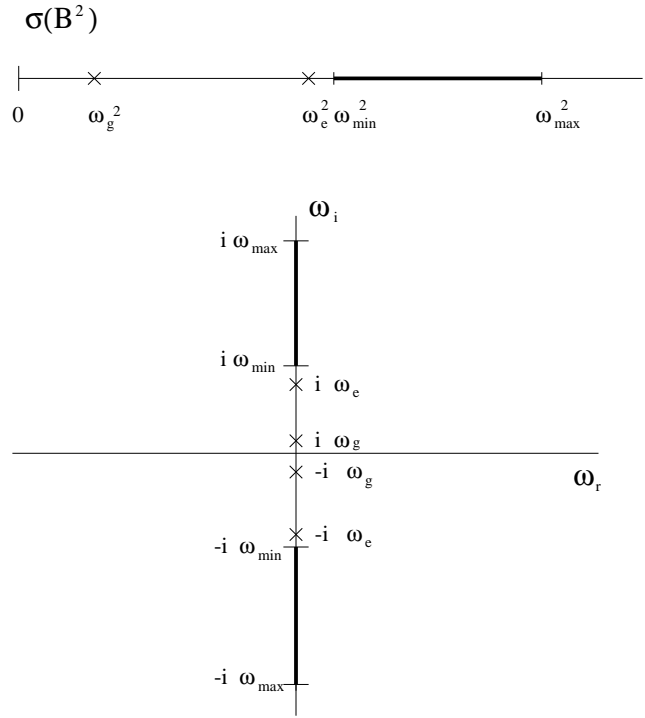,height=4.0in,width=4.2in,angle=270}}
\caption{Upper panel: schematic of spectrum of $B^2$, $\sigma(B^2)$,
 consisting of two (for $d>d_e\sim 0.515$) 
positive discrete eigenvalues ($\omega_g^2<\omega_e^2$) and 
a finite band of continuous spectrum extending from $\omega_{min}^2=1/d^2$
to $\omega_{max}^2=\sqrt{4+1/d^2}$. Lower panel: schematic of
spectrum of the kink, given by $\pm i\sigma(B)$.}
\label{fig:SGspectrum}
\end{figure}
\medskip

\begin{figure}[ptbh]
\centerline{\bf Discrete SG ground state kink and its internal modes}
\centerline{\psfig{width=6in,height=3.9in,file=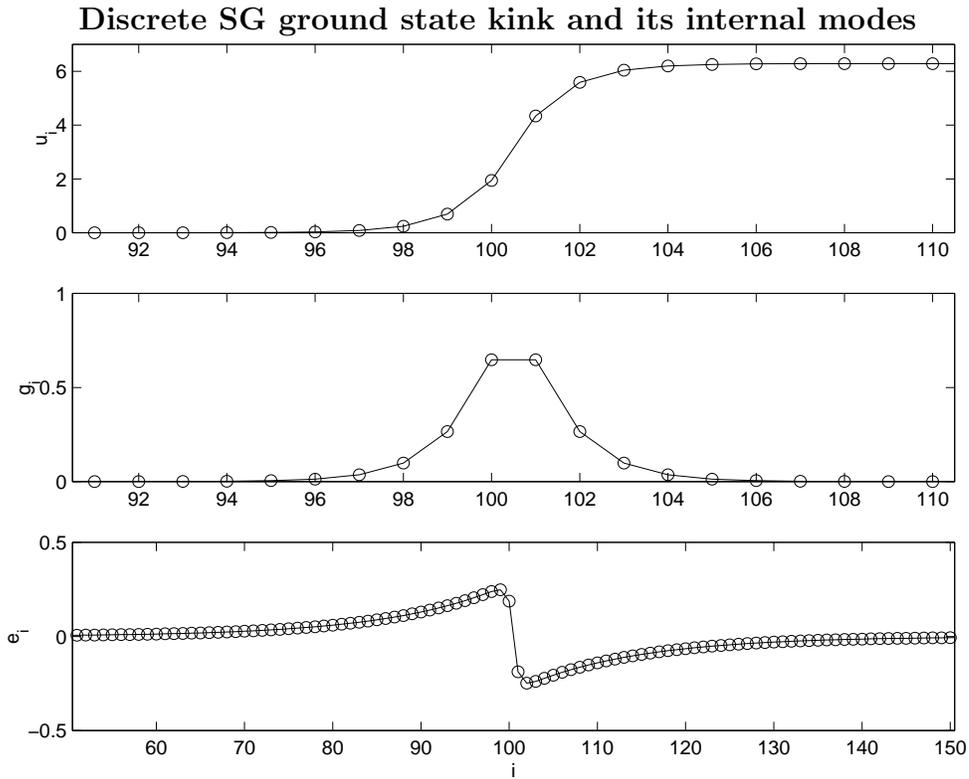}}
\caption{
 Three panels in order correspond to (a) discrete SG kink, 
 (b) spatially even Goldstone mode (c) spatially odd edge mode (present for 
 $d>d_e\sim 0.515$)}
\label{fig:SGmodes}
\end{figure}

\subsection{Discrete $\phi^4$ Model}

Consider the discrete $\phi^4$ equation:
\begin{eqnarray}
u_{i,tt}=(u_{i+1}+u_{i-1}-2 u_i)+\frac{1}{d^2}(u_i-u_i^3)
\label{dphi4}
\end{eqnarray}

In most ways the situation is quite similar to that discussed for 
 the  discrete SG. The $\phi^4$ model has low and high energy kinks.
The low energy kink is centered between lattice sites while the high
energy kink is centered at a lattice site. As $d$ increases the energy
difference, the PN barrier, tends to zero, and finally, in the scaled
limit $u(i;d^{-1})=K_i$, with $i/d\equiv x$ fixed, converges ot the 
continuum $\phi^4$ kink, 
\be K_{\phi^4}(x)\ =\ \tanh(x/\sqrt{2}).\nn\ee

Most qualitative properties of the spectrum of the linearization   
 about the $\phi^4$ kink are analogous 
to those in the SG case. 
The single key difference can be traced to the spectrum associated with 
 the continuum case. In addition to the zero (Goldstone) mode,
 the spectrum of the operators $B^2$ has internal  
 mode of odd parity with eigenfrequency $\omega_s^2= 3/(4d^2)$.
 Therefore, the spectrum of the kink has
purely imaginary internal modes associated with the energies
  $\omega_s=\pm i\sqrt{3}/(2d)$. These modes are called 
 {\it shape modes}. 

Turning now to the discrete $\phi^4$ model
we see that $B^2$ for the discrete case has the following 
properties:

\begin{itemize}
\item The dispersion relation is $\omega^2= 2/d^2+ 4\sin^2(k/2)$,
and therefore $B^2$ has continuous spectrum extending from $2/d^2$
to $2/d^2+4$. 

\item $B^2$ has a positive internal Goldstone mode 
 with energy 
 $\omega_g^2$, so that $B^2g=\omega_g^2 g$.
  This mode, which is traceable to the zero mode of the continuum model,
  is spatially even and $\omega_g$ is exponentially small in $d$ \cite{JKK}.

\item $B^2$ has an internal odd {\it shape mode} 
 with energy $\omega_s^2$, so 
that  $B^2s=\omega_s^2 s$. This mode, which is traceable to the internal
shape mode of the continuum $\phi^4$ model, is spatially odd. 

\item For $d>d_e$ ($d_e\sim 0.82$) $B^2$ has an internal {\it edge mode} 
  which is even and  with
 corresponding energy $\omega_e^2$, with $B^2e=\omega_e^2e$,
which emerges from the continuous spectrum. For large $d$ we have \cite{KPCP,KJ}
\be {\omega_e}^2 \sim 2d^{-2}-{2\over{15^2}} d^{-6}.\nn\ee
 
\end{itemize}

It follows that the spectrum of the discrete $\phi^4$ kink consists of 
 pair of two or three complex conjugate pairs of internal modes at energies
$\pm i\omega_g$, $\pm i\omega_s$, and for $d>d_e$, $\pm i\omega_e$.

\begin{figure}[t]
\centerline{\bf Spectrum of discrete $\phi^4$ ground state kink}
\centerline{\psfig{file=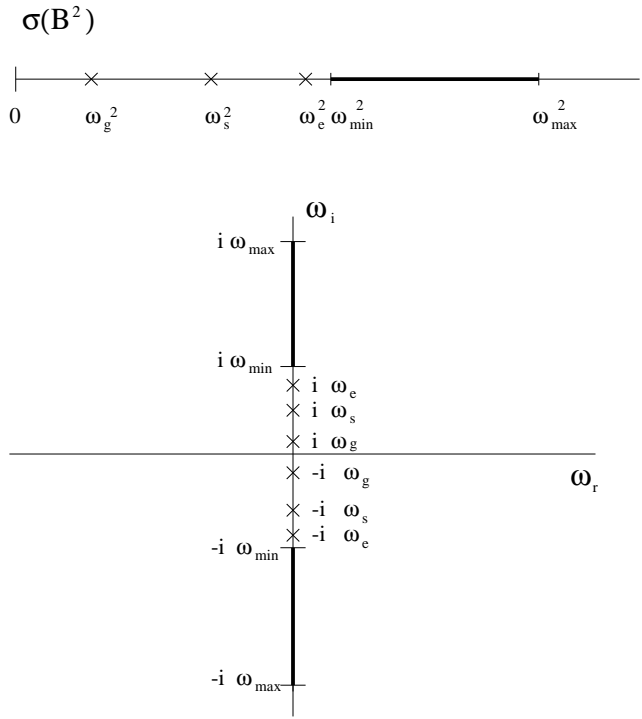,height=4.0in,width=4.2in,angle=270}}
\caption{Upper panel: schematic of spectrum of $B^2$, $\sigma(B^2)$,
 consisting of three (for $d>d_e\sim 0.82$)
positive discrete eigenvalues ($\omega_g^2<\omega_s^2<\omega_e^2$) and
a finite band of continuous spectrum extending from $\omega_{min}^2=2/d^2$
to $\omega_{max}^2=\sqrt{4+2/d^2}$. Lower panel: schematic of
spectrum of the kink, given by $\pm i\sigma(B)$.}
\label{fig:phi4spec}
\end{figure}
\bigskip

\begin{figure}[t]
\centerline{\bf Discrete $\phi^4$ ground state kink and its internal modes}
\centerline{\psfig{file=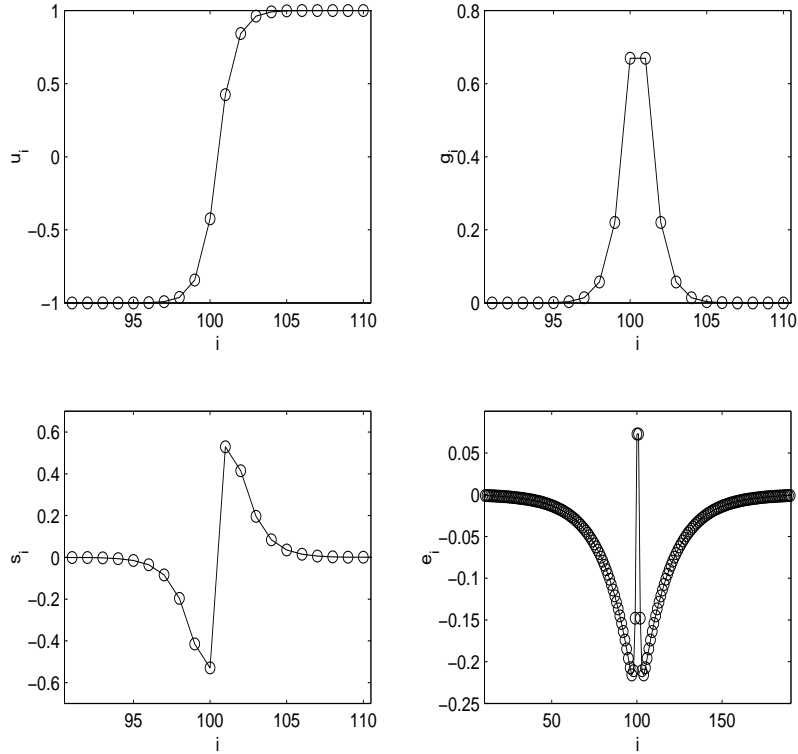,height=4.0in,width=4.2in,angle=0}}
\caption{ Top left panel as in figure one but for $\phi^4$ and
$d=0.9$. Top right shows the $g$ eigenfunction, bottom left
the $s_i$ and right the $e_i$.}
\label{fig:phi4modes}
\end{figure}

In the next section we shall see that whether or not certain integer linear 
combinations of the internal mode frequencies land in the continuous spectrum
determines the large time asymptotic behavior.

\section{Resonances and radiation damping}
\setcounter{equation}{0}

That we can Lorentz boost a static kink and obtain a traveling kink
 on the spatial continuum (section 2) motivates the following question:

{\it What happens if we attempt to cause a kink to
propagate through the lattice?}

 The  numerical and formal asymptotic
 investigation was initiated in
 \cite{PK},\cite{PA}. 
 For example, choose as initial conditions for the discrete sine Gordon 
 system the
 state $u_i(t=0)=K(i/\sqrt{1-v^2})$, where $K(x)$ is the continuum kink and
$0<|v|<1$.
 Dispersive radiation plays
an important role in the dynamics.  
 The kink moves approximately as particle under the influence of 
  the sinusoidal (Peierls-Nabarro) potential. 
 As the kink propagates, its velocity alternately increases and decreases,   
 depending on the location of its center of mass relative to the peaks and 
troughs of the potential. 
This oscillation leads to a resonance with the continuous spectrum
\cite{PK,BSW} and a transfer of energy from the propagating kink
to radiation modes. Eventually, the kink has slowed so that its energy 
no longer exceeds the PN barrier and then executes a damped 
 oscillation about a fixed lattice
site to which it is pinned. It is this latter process that we study here.

The goal of this study is to, by analytic and numerical means, 
\begin{itemize}
\item identify what the asymptotic state of the system is in different regimes 
 of the discretization parameter $d$,
\item find the rate of approach to this state.
\end{itemize}

We begin in the setting of the discrete $\phi^4$ model, (\ref{dphi4}). 
 This case
 is a more complicated than  discrete SG;  discrete $\phi^4$ has 
either two or three internal modes, while discrete SG which has at most two. 
After a detailed treatment of the $\phi^4$ case, we'll give an outline of the
analogous results for SG. 

We begin with a solution  
which is a small perturbation about a low energy (ground state) kink.
This corresponds to solving the 
 initial value problem for (\ref{dphi4}) with 
 initial conditions:
\be u_i(0)=K_i + \ve v_{0i},\ \D_tu_i(0)= \ve v_{1i},\nn\ee
where $\ve$ a small paramater. We then view the solution as
\be u_i(t)\ \equiv\ K_i\ + \ve v_i(t),\nn\ee
where $v_i(t)$ denotes the perturbation about the kink. Equation 
 (\ref{dphi4})  implies an evolution equation for $v_i$:
$\left(\D_t^2 + B^2\right)v\ =\ ...$ and since
the data for $v_i$ is small, it is  
  natural to expand $v_i$ 
 in terms of the modes of the linearization about the kink. Recall that
 the operator $B^2$ is positive and self-adjoint, and has, 
 depending on the range of $d's$ considered,
two or three internal modes $g$, $s$, and $e$ (if $d>d_e$):
\ba && B^2g_i = \omega_g^2g_i,\ B^2s_i=\omega_s^2s_i,\ B^2e_i=\omega_e^2e_i\nn\\
  &&       0<\omega_g^2<\omega_s^2<\omega_e^2< 4+2d^{-2},\nn
\ea
where
\be B^2\psi_i \equiv -\delta^2\psi_i + d^{-2}(3K_i^2-1)\psi_i.\nn\ee
We assume these internal modes to be normalized so that the 
 set $\{g,s,e\}$ is orthonormal in $l^2(\Z)$.
We also introduce the operator which projects onto the orthogonal complement
of the subspace of internal modes:
\be P_c\psi\equiv \psi-\lan g,\psi\ran g-\lan s,\psi\ran s-
 \lan e,\psi\ran e\label{eq:Pc}\ee
$P_c$ is the projection onto the continuous spectral part of $B^2$ (phonon 
or radiation modes)
and therefore the solutions of $(\D_t^2 + B^2) u=0$ with data in the range
of $P_c$ are expected to decay dispersively as $t\to\pm\infty$.

 We  
expand the perturbation $\ve v_i$ in terms of the internal mode subspace and
its orthogonal complement and thus
\begin{eqnarray}
u_i(t)=K_i+ \ve a(t) g_i+ \ve b(t) s_i + \ve c(t) e_i + \ve^2\eta_i(t),
\label{eq3}
\end{eqnarray}
with
\ba  && \lan g,\eta(t)\ran=\lan s,\eta(t)\ran=\lan e,\eta(t)\ran=0,\nn\\
     && P_c\eta = \eta\nn
\ea

Substitution of  (\ref{eq3}) into (\ref{dphi4}) yields
\ba
&& (a_{tt}+{\omega_g}^2 a) g_i + (b_{tt}+ {\omega_s}^2 b) s_i
+(c_{tt}+{\omega_e}^2 c) e_i + \ve \D_t^2\eta_i\nn\\ 
&& =\  -\ve B^2\eta - d^{-2}(3\ve K_i M_i^2 + \ve^2 M_i^3+ 6\ve^2 K_i \eta M_i
+  3\ve^3 M_i^2 \eta_i
+{\cal O}(\ve^4))
\label{eq7}
\end{eqnarray}
where: 
\be M_i\equiv a(t) g_i+ b(t) s_i + c(t) e_i.\label{eq:Mdef}\ee
Notice that from here on for simplicity and compactness we drop
the spatial indices $i$.

We next project (\ref{eq7}) onto the internal modes $g,s$ and $e$, as well
as on the range of $P_c$. This yields the following coupled system of 
four equations for the three internal mode amplitudes and the 
 continuous spectral part:  
\ba
\D_t^2a+{\omega_g}^2 a&=&-\frac{1}{d^2}\left(3\ve\lan g,KM^2\ran + 
 \ve^2\lan g,M^3\ran
+6\ve^2\lan g,K M \eta\ran +6\ve^3\lan g,M^2 \eta\ran\right) 
 + {\cal O}(\ve^4)\nn\\
&& \label{eq8}
\\
\D_t^2b+{\omega_s}^2 b&=&-\frac{1}{d^2}\left(3\ve \lan s,KM^2\ran + 
 \ve^2\lan s,M^3\ran
+6\ve^2\lan s,KM \eta\ran +6\ve^3\lan s,M^2 \eta\ran \right) + {\cal O}(\ve^4) 
\nn\\ && \label{eq9}
\\
\D_t^2c+{\omega_e}^2 c&=&-\frac{1}{d^2}\left(3\ve \lan e,KM^2\ran +
 \ve^2\lan e,M^3\ran
+6\ve^2\lan e,K M \eta\ran +6\ve^3\lan e,M^2 \eta \ran\right)+ {\cal O}(\ve^4))\nn\\  && \label{eq10}\\
\D_t^2\eta+B^2 \eta&=&-\frac{1}{d^2} P_c\left(\ve M^3 +3 K M^2+ 
 {\cal O}(\ve \eta M) + {\cal O}(\ve ^4\eta^3)
 \right).
\label{eq11}
\end{eqnarray}
The initial data for $a(t), b(t), c(t)$ is:
\ba a(0)&=&\lan g,v_0\ran,\ \D_ta(0)=\lan g,v_1\ran\nn\\
    b(0)&=&\lan s,v_0\ran,\ \D_tb(0)=\lan s,v_1\ran\nn\\
    c(0)&=&\lan e,v_0\ran,\  \D_tb(0)=\lan e,v_1\ran\nn\\
 \eta(0)&=& P_cv_0,\ \ \D_t\eta(0)=P_cv_1\label{eq:data}\ea
For simplicity we shall assume:
\be \eta(0)=0,\ \D_t\eta(0)=0\label{eq:vdata}\ee
corresponding to perturbations of the kink which only excite the internal 
modes. The general case can be treated as well; see \cite{SWinvent}.

The system (\ref{eq8}-\ref{eq11}) may be viewed as finite dimensional
Hamiltonian system  governing three oscillators with amplitudes
 $a,b$ and $c$ and  natural frequencies
$\pm\omega_g,\pm\omega_s,\pm\omega_e$, coupled by nonlinearity to an 
infinite dimensional Hamiltonian wave equation for a field 
 $\eta$. Systems of this type 
have been analyzed rigorously and the behavior
of their solutions determined on short, intermediate and 
 infinite time scales \cite{SWjsp,SWinvent}. The analysis
we present uses and extends the methods of these works.
Roughly speaking we transform the system 
 (\ref{eq8})-(\ref{eq11}) into an equivalent dynamical system, which
is a perturbation of a {\it normal form} for which one obtains information on:
 (i) what the asymptotic state of the system is,
and
 (ii) how this asymptotic state is approached.

\medskip

\section{ The dispersive normal form}  
\setcounter{equation}{0}

Our goal in this section is to derive  the normal forms
  which  give detailed information on 
the dynamics in a neighborhood of the ground state
  static kink of SG and $\phi^4$.  
 In particular, the normal forms anticipate the existence of time-periodic
solutions. These predictions are upheld by the rigorous results of
section 5. 
 In section 6, the  normal form analysis is then 
 joined with the existence theory
of section 5 and numerical simulations to fill out the picture of what
 the dynamics are in a neighborhood of the kink.
 
\subsection{The normal form for discrete $\phi^4$}
At this stage, we have in (\ref{eq8}-\ref{eq11}) a formulation of the
 discrete $\phi^4$ dynamical as a system governing the discrete internal modes
interacting, due to nonlinearity, with a system governing dispersive waves 
 (phonons). 
Our goal in this section is to present and implement 
 a method, based on \cite{SWjsp,SWinvent}, leading to a reformulation
 of the coupled discrete-continuum mode system (\ref{eq8}-\ref{eq11})
as a perturbed {\it dispersive normal form} in which the nature of the 
 energy transfer
 among modes  is made explicit.

Ideally, one could solve the equation for the dispersive part, $\eta$,
explicitly in terms of the discrete mode amplitudes: $a,b$, and $c$
and then substitute the result into the mode amplitude equations to get a 
closed system of equations for $a,b$, and $c$. This then could be used 
in the equation for $\eta$ to determine its behavior. 
 Due to the system's nonlinearity, one can't expect to solve for the 
 exact behavior of $\eta$  in terms of $a,b$ and $c$. Instead we solve 
 perturbatively in $\ve$. Let
\be \eta \equiv \sum_{j=0}^{\infty} \ve^j \eta^{(j)}. \nn\ee
Then,
\be \D_t^2\eta^{(0)} + B^2\eta^{(0)} = -3d^{-2} K M^2\label{eq:eta1}.\ee
The function 
$\eta$ captures the leading order radiative response, to the
internal mode excitations.

Note that 
 $\eta^{(0)}$ is the solution of a forced wave equation. The forcing term 
 in (\ref{eq:eta1}) involves $M^2$,
 which  by (\ref{eq:Mdef}), contains squares and cubes of $a,b$ and $c$. 
 Also,
note that since the nonlinear coupling terms on the 
 right hand sides of equations 
 (\ref{eq8}-\ref{eq10}) are small,  
  $a(t), b(t)$ and $c(t)$  
 are slow modulations of the exponentials $e^{\pm i\omega_gt}$,
 $e^{\pm i\omega_st}$ and $e^{\pm i\omega_et}$.
The forcing on the right hand side of (\ref{eq:eta1}) will be resonant if 
nonlinearity, when acting on $a,b$ and $c$, generates frequencies which 
lie in the spectrum of the operator $B$. Since $B$ has a band
  of continuous 
spectrum, it can easily happen that 
integer linear combinations of the internal mode frequencies can 
lie in the continuous spectrum of $B$. 

Figure 6  in section 4 displays the locations
of the frequencies $\omega_g, \omega_s$ and $\omega_e$, their multiples and
certain integer linear combinations relative to the continuous (phonon)
 spectrum. 
Figure 7 is the analogous plot for discrete SG.

The key calculation is to compute
the effect of such resonances which result in the transfer of 
 energy from the discrete to continuum modes.  

Let
\ba
a(t)&=&A(t) \exp(i \omega_g t) + \bar{A}(t) \exp(-i \omega_g t),
\label{eq14}
\\
b(t)&=&B(t) \exp(i \omega_s t) + \bar{B}(t) \exp(-i \omega_s t),
\label{eq15}
\\
c(t)&=&C(t) \exp(i \omega_e t) + \bar{C}(t) \exp(-i \omega_e t),
\label{eq16}
\ea
We further impose the constraints:
\ba A_t \exp(i\omega_g t) + \bar{A}_t \exp(-i \omega_g t) &=& 0\nn\\
    B_t \exp(i\omega_s t) + \bar{B}_t \exp(-i \omega_s t) &=& 0\nn\\
    C_t \exp(i\omega_e t) + \bar{C}_t \exp(-i \omega_e t) &=& 0.\nn
\ea

and obtain:
\ba
A_t &=& (2i\omega_g)^{-1}e^{-i\omega_g t} F_1
\label{eq15a}
\\
B_t &=& (2i\omega_s)^{-1}e^{-i\omega_s t} F_2
\label{eq16a}
\\
C_t &=& (2i\omega_e)^{-1}e^{-i\omega_e t} F_3
\label{eq17}
\ea
where $F_i=F_{i1}+F_{i2}$ (i=1,2,3). 
The $3\times2$ matrix $F_{ij}$ is given by
\ba
F &=& -d^{-2}\left [
\begin{array}{rr}
3\ve \lan g,K M^2\ran +\ve^2 \lan g,M^3\ran & 
6\ve^2\lan g,K M \eta\ran + 3\ve^3 \lan g,M^2 \eta\ran\\
3\ve\lan s,K M^2\ran + \ve^2\lan s,M^3\ran & 
6\ve^2\lan s,K M \eta\ran + 3\ve^3\lan s,M^2 \eta\ran\\
3\ve\lan e,K M^2\ran+\ve^2\lan e,M^3\ran & 
6\ve^2\lan e,K M \eta\ran + 3\ve^3\lan e,M^2\eta\ran\\
\end{array}
\right]
\nn\\ &&\label{eq:Fmatrix}\ea
 
We first solve (\ref{eq:eta1}) and obtain the expression 
for $\eta^{(0)}$ in terms of
$a(t),b(t)$ and $c(t)$ or equivalently $A(t), B(t)$ and $C(t)$
\begin{eqnarray}
\eta^{(0)}=-3d^{-2}\int_0^t {\sin(B(t-\tau))\over B}\ P_c K M^2 d\tau
\label{eq13}
\end{eqnarray}
Recall that the dependence on the internal mode amplitudes is through
$M$, defined in (\ref{eq:Mdef}).
Substitution of (\ref{eq13}) into  equations (\ref{eq15a} 
-\ref{eq17}) yields a closed system for the internal mode amplitudes
through order $\ve^3$. The key terms in this equation come from 
certain resonances and our goal now is to show how they arise. 

The first column of terms in $F$ involves the interactions among 
discrete modes. They do not generate frequencies contributing to any 
 resonant forcing. 
The second column contains terms which  
 couple the discrete bound state part
 to the continuum radiation modes. 
 In order to identify all the 
resonances one has to explicitly expand out the $F_{i2}$ terms
and to use equation (\ref{eq13}). We do not carry out the detailed computations
in all detail here. Rather, we illustrate the key ideas and methodology  
 by considering prototypical terms. A complete and rigorous implementation 
of these ideas in another nonlinear wave context is presented in  
\cite{SWinvent}.

We focus on the ${\cal O}(\ve^2)$ term in $F_{12}$ 
and  {\it some} of its contributions 
 to the equation for $A(t)$. In particular we shall consider 
all resonant contributions and 
 a sample
  nonresonant contribution. By equation (\ref{eq15a}) we must then consider
the expression:
\be 3i\ve^2d^{-2}\omega_g^{-1}e^{-i\omega_gt}\lan Kg,M\eta^{(0)}\ran.
\label{eq:F12ve2}\ee
Consider $\eta^{(0)}$.
\ba
&& \eta^{(0)} = -3d^{-2}\int_0^t {\sin(B(t-\tau))\over B}\   P_c  K M^2 
 d\tau\ =\ 
  -{3\over 2iB} d^{-2} \int_0^t e^{iB(t-\tau)}\ P_c K M^2 d\tau\ +\ ...\nn\\
 &=& {3i\over 2B}d^{-2}e^{iBt} \int_0^t   e^{-iB\tau}\ P_c Kg^2 a^2(\tau) 
     d\tau\ +\ ...\nn\\
 &=& {3i\over 2B}e^{iBt} \int_0^t e^{-iB\tau} 
     \left(A^2(\tau)e^{2i\omega_g\tau}+2|A(\tau)|^2+
     \bar A^2(\tau)e^{-2i\omega_g\tau}\right)\ P_c Kg^2 \ d\tau +...\nn\\
 &=& {3i\over 2B}e^{iBt}\int_0^t A^2(\tau)e^{-i(B-2\omega_g)\tau}\ 
 P_cKg^2 d\tau +
 {3i\over 2B}e^{iBt}\int_0^t e^{-i(B+2\omega_g)\tau} 
      \bar A^2(\tau)\ P_cKg^2 d\tau +...\nn\\
&\equiv& \eta^{(0)}_{res} + \eta^{(0)}_{nr} + ...
\label{eta1res}
\ea
We have only kept two terms in the above calculation: the only term leading
to a resonant contribution in the $A$ equation and one (of many) nonresonant 
terms.

We wish to expand $\eta^{(0)}$ using the following formula, which follows
 by straightforward integration by parts: 
\ba && e^{iBt}\int_0^t e^{-i(B-\zeta)\tau} P_c \alpha(\tau) d\tau\nn\\
  &=& ie^{iBt} (B-\zeta)^{-1}P_c e^{i\zeta t} \alpha(\tau)-
 ie^{iBt} (B-\zeta)^{-1}P_c \alpha(0)\nn\\ 
 && -ie^{iBt} \int_0^t e^{-i(B-\zeta)\tau} (B-\zeta)^{-1}P_c\ \D_\tau\alpha(\tau) d\tau
 \label{eq:ibpnonres}\ea
with $\alpha(\tau) = A^2(\tau)\ P_cKg^2$, for example, and 
 that $\tau$-derivatives of $A$ are of order $\ve$. For the term 
 $\eta^{(0)}_{res}$, $\zeta=2\omega_g$, which for $d$ in the range 
$[0.54, 0.6364]$
 lies in the continuous spectrum; see Table 2 and  Figure 6 
  below.
  Such resonant terms are of paramount interest
and govern energy transfer. To treat these resonant terms, an
 appropriate modification of 
 (\ref{eq:ibpnonres}) is required. We use the following:

\noindent  Let $\kappa=sgn(t)$, which is equal to
  $+1$ if $t>0$ and $-1$ for $t<0$.
\ba 
&& e^{iBt}\int_0^t e^{-i(B-\zeta)\tau}P_c\  \alpha(\tau) d\tau =\nn\\ 
&& i e^{iBt}  (B-\zeta+i\kappa 0)^{-1}P_c\ e^{i\zeta t} \alpha(\tau)
-ie^{iBt}  (B-\zeta+i\kappa 0)^{-1}P_c\ \alpha(0) \nn\\
&& -ie^{iBt} \int_0^t e^{-i(B-\zeta)\tau}(B-\zeta+i\kappa 0)^{-1}P_c\ 
 \D_\tau\alpha(\tau) d\tau,
 \label{eq:ibpres}\ea
where 
\be (B-\zeta+i\kappa 0)^{-1}=\lim_{\epsilon\downarrow0} 
 (B-\zeta+i\kappa\epsilon)^{-1}.\nn\ee
 
\begin{rmk}

\noindent (1) The sense in which the operator expansions (\ref{eq:ibpnonres}) 
 and (\ref{eq:ibpres}) are correct is in a distributional sense. That is,
 equality holds when multiplying both sides by a smooth function 
 with spatial support which is compact and integrating both 
 sides over all space.

\noindent (2) Formula
 (\ref{eq:ibpres}) is proved by first writing the integral on the 
 left hand side as 
\be
\int_0^t\exp(-i(B-\zeta+i\kappa\epsilon)\tau)\
  P_c\alpha(\tau)\ d\tau.
\nn\ee 
For any $\epsilon>0$  (\ref{eq:ibpnonres}) can be 
used with $\zeta$ replaced by $\zeta-i\epsilon$. We then pass to the limit as
$\epsilon\downarrow0$.

\noindent (3) The choice of regularization, $+i0\ (\kappa=1)$ for $t>0$ 
 and $-i0\ (\kappa=-1)$ for $t<0$ is 
connected with the condition that the latter two terms in 
(\ref{eq:ibpres}) consist of outgoing radiation at spatial infinity, 
 and is therefore time-decaying in an appropriate local energy sense 
 \cite{SWinvent}.
In this article, we take into account
  only the first term in (\ref{eq:ibpres}). Reference 
 \cite{SWinvent}  contains a fully detailed and rigorous treatment in a related 
context. Henceforth, we shall for simplicity assume $t>0$ and therefore
 work with the $+i0$ regularization.

\noindent (4) If $\zeta$ does not lie in the continuous spectrum the formula
(\ref{eq:ibpres}) reduces to (\ref{eq:ibpnonres}). 
\end{rmk}

We now continue the expansion of $\eta^{(0)}$ using (\ref{eq:ibpnonres}) 
to study $\eta^{(0)}_{nr}$ and (\ref{eq:ibpres}) to study 
 $\eta^{(0)}_{res}$:
\ba
\eta^{(0)}_{res} &=& 
 -{3\over2}d^{-2} B^{-1}(B-2\omega_g+i0)^{-1}A(t)^2\ P_cKg^2 + ...\nn\\
\eta^{(0)}_{nr}  &=&
  -{3\over2}d^{-2} B^{-1}(B+2\omega_g)^{-1}\bar A(t)^2\ P_cKg^2 + ...
\nn\ea
Substitution of $\eta^{(0)}=\eta^{(0)}_{res}+\eta^{(0)}_{nr}+...$ into
(\ref{eq:F12ve2}) we have:
\ba  
&&3i\ve^2d^{-2}\omega_g^{-1} e^{-i\omega_gt}\lan Kg,M\eta^{(0)}\ran\nn\\
&& = 
3i\ve^2d^{-2}\omega_g^{-1}e^{-i\omega_gt}
 \left (Ae^{i\omega_gt}+\bar Ae^{-i\omega_gt}\right) 
\lan Kg^2, \eta^{(0)}\ran +...\nn\\
&&= 
 -{9\over2}i\ve^2d^{-4}\omega_g^{-1} |A|^2A \lan Kg^2,
 B^{-1}(B-2\omega_g+i0)^{-1}\ P_c Kg^2 + ... \ran\nn\\ 
&&-{9\over2}i\ve^2d^{-4}\omega_g^{-1} \bar A^3 e^{-3i\omega_gt}\lan Kg^2,
 B^{-1}(B+2\omega_g)^{-1}\ P_c Kg^2 \ran + ...\nn\\
&&\equiv \ve^2\left( -\Gamma_{2\omega_g} + i \Lambda_{2\omega_g} \right)|A(t)|^2A(t) +
 \rho(t) \bar A^3(t)\label{eq:term1}\ea

To calculate $\Lambda$ and $\Gamma$, we apply a generalization to self-adjoint
operators of the well-known distributional identity:
\be
\lim_{\epsilon\to0}(\xi\pm i\epsilon)^{-1} = {\rm P.V.}\ \xi^{-1}
\mp i \pi \delta(\xi),\label{eq:plemelj}\ee
where $\delta(\xi)$ is the Dirac delta mass at $\xi=0$ and 
${\rm P.V.}$ denotes the principal value integral.

Therefore, using (\ref{eq:plemelj}) we obtain:
\ba
\ve^2\Gamma_{2\omega_g}&\equiv& \ve^2\Gamma(2\omega_g)= {9\pi\over 4\omega_g^2}\ve^2d^{-4}\ 
 \lan Kg^2,\delta(B-2\omega_g) Kg^2\ran, \label{eq:Gamma}\\
\ve^2\Lambda_{2\omega_g}&\equiv& \ve^2\Lambda(2\omega_g)= -{9\over2} \ve^2 d^{-4} 
 \lan Kg^2, P.V.(B-2\omega_g) P_c Kg^2\ran.
\label{eq:Lambda}\ea

\begin{rmk}
\noindent (1) Had we done this calculation for the $B$ (respectively, $C$) 
 equation,
we'd  have obtained as a coefficient of $|B|^2B$ (respectively, $|C|^2C$)
 the quantity $\ve^2(-\Gamma_{2\omega_s}+i\Lambda_{2\omega_s})$ 
(respectively, $\ve^2(-\Gamma_{2\omega_e}+i\Lambda_{2\omega_e})$) 
\noindent (2) If $\zeta\in\sigma_{cont}(B)$,
 then $\Gamma_\zeta$ is always nonnegative and, generically,
 is strictly positive. It is the analogue of {\it Fermi's
golden rule}
wwhich arises in the context of the theory of spontaneous emission 
\cite{LL},\cite{GAFA}.
 Apart from a positive constant prefactor, $\Gamma_\zeta$ 
 is the square of the Fourier
transform of $Kg^2$ relative to the continuous spectral part of 
$B$ evaluated at $\zeta$.
\end{rmk}

The above calculations 
 yield the following information on the structure 
of the equation for $A(t)$: 

\be
A_t = \ve^2\left(-\Gamma_{2\omega_g} +i\Lambda_{2\omega_g} \right)|A|^2A + ....
\label{eq:A1}\ee

Extensive calculations, of which the above are representative, 
yield a system of the form:
\ba 
A_t &=&
 \ve^2\left(\alpha_1|A|^2 +\alpha_2|B|^2+\alpha_3|C|^2\right) A\nn\\
&&\ve^4\left(\alpha_4|A|^4 + \alpha_5|A|^2|B|^2+
  \alpha_6|B|^4+\alpha_7|B|^2|C|^2+ \alpha_8|C|^4\right) A\nn\\
&+& \ve\Xi_A(t,A,B,C;\ve)\nn\ea
\ba 
B_t &=&
\ve^2\left(\beta_|A|^2 +\beta_2|B|^2+\beta_3|C|^2\right) B\nn\\
&&\ve^4\left(\beta_4|A|^4 + \beta_5|A|^2|B|^2+
  \beta_6|B|^4+\beta_7|B|^2|C|^2+\beta_8|C|^4\right) B\nn\\
&+& \ve\Xi_B(t,A,B,C;\ve)\nn\ea
\ba 
C_t &=&
 \ve^2\left(\gamma_1|A|^2 +\gamma_2|B|^2+\gamma_3|C|^2\right) C\nn\\
&&\ve^4\left(\gamma_4|A|^4 + \gamma_5|A|^2|B|^2+
  \gamma_6|B|^4+\gamma_7|B|^2|C|^2+ \gamma_8|C|^4\right) C\nn\\
&+& \ve\Xi_C(t,A,B,C;\ve)\label{eq:ABCt}\ea

The coefficients $\alpha_j$, $\beta_j$ and $\gamma_j$ are, in 
 general, complex  numbers. 
The terms $\Xi_A$, $\Xi_A$ and 
 $\Xi_A$  involve
 bounded and  oscillatory complex exponentials in $t$ multiplying monomials
in $A,B,C$ of cubic or higher degree. 
Also, in order to obtain the fifth degree terms it is necessary to construct
$\eta$ through second order in $\ve$ (and therefore the continuous spectral 
part of the perturbation about the kink, $\ve^2\eta$, through $\ve^4$.
 Coupling to $\eta$ is neglected as
is the dynamical equation for $\eta$.
 The preceding calculation
gives $\alpha_1=-\Gamma_g+i\Lambda_g$ plus a further term contributed 
by the ${\cal O}(\ve^2)$ entry of $F_{11}$ in (\ref{eq:Fmatrix}). 
An exhaustive tabulation of all coefficients $\alpha_j,\beta_j, \gamma_j$
would be, to put it midly, a very lengthy exercise. Since the principal 
effect we seek to illuminate is that of nonlinear resonant coupling
on the internal mode amplitudes $|A|, |B|$ and $|C|$, we only 
tabulate those coefficients up to the order considered which may
play a role. 

 For this it is convenient to introduce the
notation:
 \be G(\zeta)=  P_c B^{-1}(B-\zeta+i0)^{-1} P_c \label{eq:Rzeta}\ee
{\footnotesize


\centerline{\bf Table 1A: Principal $\phi^4$ g-mode coefficients}
\begin{center}
\begin{tabular}{|c|c|} \hline
 Term  &   Coefficient ($\alpha_j$)  \\ \hline
$|A|^2A$       &  $-{9\over2}id^{-4}\omega_g^{-1}
\lan Kg^2,G(2\omega_g) Kg^2\ran  $  \\ \hline
$|B|^2A$      &  $-9id^{-4}\omega_g^{-1}
                 \lan Kgs,G(\omega_g+\omega_s)Kgs\ran$\\ \hline
$|C|^2A$      & $-9id^{-4}\omega_g^{-1}
                 \lan Kge,G(\omega_g+\omega_e)Kge\ran$\\ \hline
$|A|^4A$      & $-{3\over4}id^{-4}\omega_g^{-1}
                 \lan g^3,G(3\omega_g)g^3\ran$\\ \hline
$|B|^4A$      & $-{9\over4}id^{-4}\omega_g^{-1}
                 \lan gs^2,G(\omega_g+2\omega_s)gs^2\ran$\\ \hline
$|C|^4A$      & $-{9\over4}id^{-4}\omega_g^{-1}
                 \lan ge^2,G(\omega_g+2\omega_e)ge^2\ran$\\ \hline
$|A|^2|B|^2A$ & $-{9\over2}id^{-4}\omega_g^{-1}
                 \lan g^2s,G(2\omega_g+\omega_s)g^2s\ran$\\ \hline
$|A|^2|C|^2A$ & $-{9\over2}id^{-4}\omega_g^{-1}
                 \lan g^2e,G(2\omega_g+\omega_e)g^2e\ran$\\ \hline
$|B|^2|C|^2A$ & $-9id^{-4}\omega_g^{-1}
                 \lan gse,G(\omega_g+\omega_s+\omega_e)gse\ran$\\ \hline
\end{tabular}
\end{center}


\centerline{\bf Table 1B: Principal $\phi^4$ s-mode coefficients}
\begin{center}
\begin{tabular}{|c|c|} \hline
 Term  &   Coefficient ($\beta_j$) \\ \hline
$|B|^2B$       &  $-{9\over2}id^{-4}\omega_s^{-1}
\lan Ks^2,G(2\omega_s) Ks^2\ran $  \\ \hline
$|A|^2B$      &  $-9id^{-4}\omega_s^{-1}
                 \lan Kgs,G(\omega_g+\omega_s)Kgs\ran$\\ \hline
$|C|^2B$      & $-9id^{-4}\omega_s^{-1}
                \lan Kse,G(\omega_s+\omega_e)Kse\ran$\\ \hline
$|B|^4B$      & $-{3\over4}id^{-4}\omega_s^{-1}
                \lan s^3,G(3\omega_s)s^3\ran$\\ \hline
$|A|^4B$      & $-{9\over4}id^{-4}\omega_s^{-1}
                \lan sg^2,G(2\omega_g+\omega_s)sg^2\ran$\\ \hline
$|C|^4B$      & $-{9\over4}id^{-4}\omega_s^{-1}
                \lan se^2,G(\omega_s+2\omega_e)se^2\ran$\\ \hline
$|A|^2|B|^2B$ & $-{9\over2}id^{-4}\omega_s^{-1}
                \lan gs^2,G(\omega_g+2\omega_s)s^2g\ran$\\ \hline
$|A|^2|B|^2B$ & $-{9\over2}id^{-4}\omega_s^{-1}
                \lan gs^2,G(2\omega_s-\omega_g)s^2g\ran$\\ \hline
$|B|^2|C|^2B$ & $-{9\over2}id^{-4}\omega_s^{-1}
                \lan s^2e,G(2\omega_s+\omega_e)s^2e\ran$\\ \hline
$|A|^2|C|^2B$ & $-9id^{-4}\omega_s^{-1}
                \lan gse,G(\omega_s+\omega_e-\omega_g)gse\ran$\\ \hline
$|A|^2|C|^2B$ & $-9id^{-4}\omega_s^{-1}
                \lan gse,G(\omega_g+\omega_s+\omega_e)gse\ran$\\ \hline
\end{tabular}
\end{center}


\centerline{\bf Table 1C: Principal $\phi^4$ e-mode coefficients}
\begin{center}
\begin{tabular}{|c|c|} \hline
 Term  &   Coefficient ($\gamma_j$) \\ \hline
$|C|^2C$       &  $-{9\over2}id^{-4}\omega_e^{-1}
\lan Ke^2,G(2\omega_e) Ke^2\ran $  \\ \hline
$|B|^2C$      &  $-9id^{-4}\omega_e^{-1}
                 \lan Kse,G(\omega_s+\omega_e)Kse\ran$\\ \hline
$|A|^2C$      & $-9id^{-4}\omega_e^{-1}
                \lan Kge,G(\omega_g+\omega_e)Kge\ran$\\ \hline
$|C|^4C$      & $-{3\over4}id^{-4}\omega_e^{-1}
                \lan e^3,G(3\omega_e)e^3\ran$\\ \hline
$|B|^4C$      & $-{9\over4}id^{-4}\omega_e^{-1}
                \lan s^2e,G(2\omega_s+\omega_e)s^2e\ran$\\ \hline
$|A|^4C$      & $-{9\over4}id^{-4}\omega_e^{-1}
                \lan g^2e,G(2\omega_g+\omega_e)g^2e\ran$\\ \hline
$|C|^2|B|^2C$ & $-{9\over2}id^{-4}\omega_e^{-1}
                \lan se^2,G(\omega_s+2\omega_e)se^2\ran$\\ \hline
$|C|^2|B|^2C$ & $-{9\over2}id^{-4}\omega_e^{-1}
                \lan se^2,G(2\omega_e-\omega_s)se^2\ran$\\ \hline
$|C|^2|A|^2C$ & $-{9\over2}id^{-4}\omega_e^{-1}
                \lan ge^2,G(\omega_g+2\omega_e)ge^2\ran$\\ \hline
$|C|^2|A|^2C$ & $-{9\over2}id^{-4}\omega_e^{-1}
                \lan ge^2,G(2\omega_e-\omega_g)ge^2\ran$\\ \hline
$|B|^2|A|^2C$ & $-9id^{-4}\omega_e^{-1}
                \lan gse,G(\omega_g+\omega_s+\omega_e)gse\ran$\\ \hline
$|B|^2|A|^2C$ & $-9id^{-4}\omega_e^{-1}
                \lan gse,G(\omega_s+\omega_e-\omega_g)gse\ran$\\ \hline
\end{tabular}
\end{center}
}

Before discussing the information contained in these tables, we note
that the system (\ref{eq:ABCt}) can be further simplified. 
 Using a near-identity
transformation:
\be \left(\tilde A, \tilde B, \tilde C\right)= \left(A, B, C\right) +
\ {\cal O}\left(|A|^2+|B|^2+|C|^2\right)\label{eq:nearidentity}\ee
the system (\ref{eq:ABCt}) can be transformed into a new system for
$\tilde A,  \tilde B$ and $\tilde C$ of a very similar form, but
 with the following modifications:

\begin{itemize}
\item The real parts of the coefficients are the same but the imaginary parts
may be modified.
\item  The $\ve^2\Xi$ terms are now replaced by terms of order $\ve^6$
\end{itemize}
We refer to the
  system governing $\tilde A$, $\tilde B$ and $\tilde C$,
  obtained in the manner, after neglecting the $\Xi$ terms,
as a {\it dispersive normal form}.

While complicated in its details, there is a simple way to think about this 
  normal form. We introduce the {\it internal mode powers}
\footnote{Strictly speaking, by (\ref{eq:nearidentity}) $P$, $Q$ and $R$ 
are only approximately equal, respectively,
  to the Goldstone, shape and edge mode powers.}:
\be
 P=|\tilde A|^2,\ Q=|\tilde B|^2,\ R\equiv |\tilde C|^2.
\label{eq:powers} 
\ee
The equations for the powers are:
\ba
P_t &=&
 2\ve^2\left(\alpha^r_1P +\alpha^r_2Q+\alpha^r_3R\right) P\nn\\
&&+2\ve^4\left(\alpha^r_4P^2 + \alpha^r_5PQ+
  \alpha^r_6Q^2+\alpha^r_7QR+ \alpha^r_8R^2\right) P\label{eq:P}\\
Q_t &=&
2\ve^2\left(\beta^r_1P +\beta^r_2Q+\beta^r_3R\right) Q\nn\\
&&+2\ve^4\left(\beta^r_4P^2 + \beta^r_5PQ+
  \beta^r_6Q^2+\beta^r_7QR +\beta^r_8R^2\right) Q\label{eq:Q}\\
R_t &=&
 2\ve^2\left(\gamma^r_1P +\gamma^r_2Q +\gamma^r_3 R\right) R\nn\\
&&+2\ve^4\left(\gamma^r_4P^2 + \gamma^r_5PQ+
  \gamma^r_6Q^2+\gamma^r_7QR+ \gamma^r_8R^2\right) R,
\label{eq:R}\ea
where $\alpha_j^r, \beta^r_j$ and $\gamma_j^r$ denote the real parts
of the coefficients $\alpha_j^r, \beta_j^r$ and $\gamma_j^r$ appearing in 
 (\ref{eq:ABCt}). The real parts do not change under the 
near identity change of variables: 
 $A\mapsto\tA, B\mapsto\tB, C\mapsto\tC$, and are therefore given by 
the real parts of the coefficients displayed in Table 1. In general,
as the  reader may note, a contribution to the above mentioned
internal mode power equations of the form $ |A|^{2 m_1} |B|^{2 m_2}
|C|^{2 m_3}$ comes from a frequency combination $m_1 \omega_g+ 
m_2 \omega_s + m_3 \omega_e$ landing in the band of continuous
spectrum  through a term
\be
(M(K,g,s,e), G(m_1 \omega_g + m_2 \omega_s + m_3 \omega_e) M(K,g,s,e))
\ee
where $M$ is the appropriate monomial combination of the relevant 
spatial parts.

As the discreteness parameter, $d$, varies the spectrum of
the kink (the continuous spectrum and
  the number and location of the internal modes) changes. As indicated in 
Figure 6, 
 for different values of $d$ various integer linear combinations
of the internal mode frequencies, the simplest of which are those appearing
 as arguments of $G(\cdot)$ in Table 1, 
  may lie in the continuous spectrum. 
\begin{figure}[ptbh]
\centerline{\bf Arithmetic of $\phi^4$ kink frequencies}
\centerline{\psfig{width=4.2in,height=3.9in,file=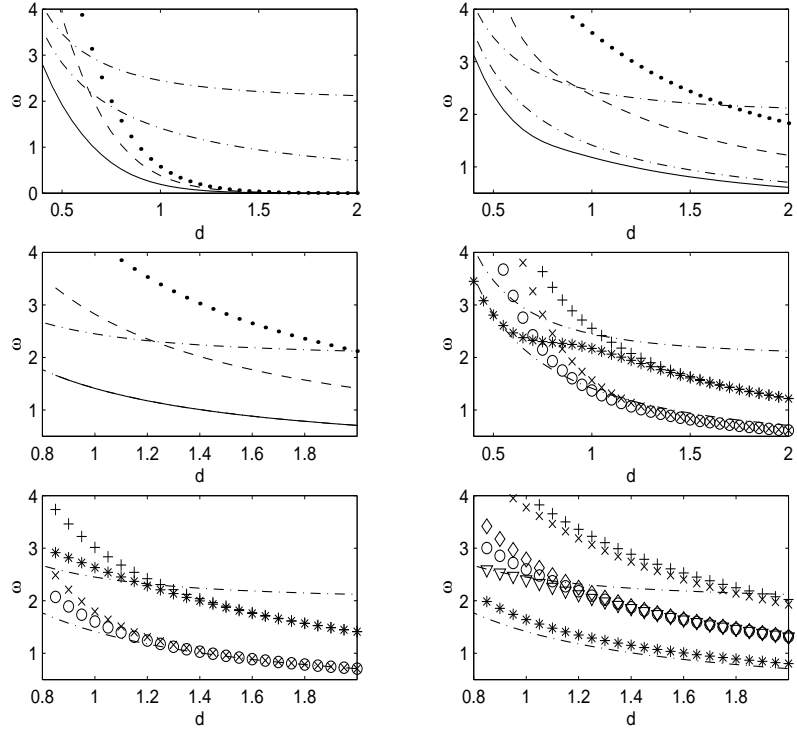}}
\label{fig:phi4combos}
  \caption{ 
 Integer linear combinations of internal mode frequencies:
In all $6$ panels, the band is indicated by dash-dotted lines.
 Top left shows the goldstone
frequency as a function of $d$ (solid) and its second (dashed)
and third (dotted) harmonics for $\phi^4$. The top right panel
follows the same sequence of symbols for the shape mode harmonics
while the left panel of the second row does so for the edge mode.
The right panel of the second row shows the combinations of
$\omega_g,\omega_s$ that can appear in the band. In 
particular: $\omega_g+\omega_s:$o$,2 \omega_g+\omega_s:$x$,
2\omega_s+\omega_g:+,2\omega_s-\omega_g:*$. Similarly
in the bottom left panel for the $g,e$ modes. 
In particular: $\omega_g+\omega_e:o,2
\omega_g+\omega_e:x,
2\omega_e+\omega_g:+,2\omega_i-\omega_g:*$. And in the case of
bottom right panel $s,e$ modes are shown but so are combinations of
all $3$ frequencies. In particular: $\omega_s+\omega_e:o,2
\omega_s+\omega_e:x,
2\omega_e+\omega_s:+,2\omega_i-\omega_s:*,\omega_s+\omega_g+\omega_e:$
diamonds$,\omega_s+\omega_e-\omega_g:$down triangles.
 }
\end{figure}

{\it How does this influence the character of the normal form?} 

 Let $\zeta$ denote one such linear combination of frequencies. By 
(\ref{eq:plemelj}):
\ba
G(\zeta)=P_c\ B^{-1}(B-\zeta)^{-1}\ P_c,\ \  \zeta\notin 
                      \sigma_{cont}(B)\nn\\
G(\zeta)=P_c\ {\rm P.V.}\ B^{-1}(B-\zeta)^{-1}\ P_c\ 
           -i{\pi\over\zeta} P_c\ \delta(B-\zeta)\ P_c,\ \  
           \zeta\in\ \sigma_{cont}(B).\label{eq:Gzeta}
\ea
Each of the coefficients of the system listed in Table 1 is a positive
 multiple of an expression of the
form $-i\lan f,G(\zeta)f\ran$, where $f$ is spatially localized. Therefore,
if $\zeta$ does not lie in the continuous spectrum of $B$, the associated 
coefficient of the system for the powers $P,Q,R$, 
 $\alpha^r=\Re\alpha, \beta^r=\Re\beta$ or $\gamma^r=\Re\gamma$,  will be zero. On the other hand, if $\zeta$ {\bf does} lie in the continuous spectrum of 
$B$, this coefficient will be of the form: 
\ba
 \Gamma_\zeta\ &\equiv&\ 
 -{\pi\over\zeta} \lan P_cf,  \delta(B-\zeta)\ P_cf \ran\nn\\
 &=&\ -{\pi\over\zeta} \left|{\cal F}_B[f](\zeta)\right|^2,  
\nn\ea
where ${\cal F}_B$ denotes the Fourier transform with respect to the 
continuous spectral part of the operator $B$. Generically, one has
$\Gamma_\zeta$ is strictly negative. Therefore, such resonances are associated
with nonlinear damping of energy in the internal modes.
  This does not contradict
the Hamiltonian character of the equations of motion. The $\tA,\tB,\tC$ 
 system is
coupled  to the dispersive system governing $\eta$; 
 damping of the discrete mode
 amplitudes implies a transfer of energy from the discrete to continuum modes.
 The information contained in Figure 6 enables us to
  determine, for each $d$, which
combinations of harmonics appear in the phonon band. 
 Then  Tables 1A, 1B and 
 1C, together with (\ref{eq:P}-\ref{eq:R}) give us the precise form of the 
internal mode power equations, from which we can ascertain 
 the detailed behavior
of solutions in a neighborhood of the static kink.

In the Table 2 (below) we present the form of the internal mode
 power equations  for 
different ranges of the parameter, $d$. 
There are numerous changes in the form of these equations as $d$ varies,
so for clarity, we indicate only the key transitions. These key transitions
occur across values of $d$ where there is a topological change in the 
phase portrait of the system governing the internal mode powers. 
Such changes are found due to a change in the nature of the set of
equilibria, {\it e.g.} going from a system with one line of equilibria 
to one where there are two lines of equilibria,
  or due to a change in the number of internal 
modes (jump in the dimensionality of the phase portrait). The latter
transition occurs at $d=d_e,\ d_e\sim 0.82$, when a point
  eigenvalue emerges
from the edge of the continuous spectrum and appears as a third internal
 mode, $e$, an  {\it edge mode}, with corresponding frequency 
 $\omega_e$.

\medskip

{\small
\centerline{\bf Table 2:  $\phi^4$ internal mode power equations}
}
{\footnotesize
\begin{center}
\begin{tabular}{|l|l|l|} \hline
Regimes of $d$& Resonances &  System Form \\ \hline
$I: d <0.5398$  & $\{2 \omega_s-\omega_g, \dots \}$ &
       $P_t=0,\ Q_t=-\ve^4\{PQ^2\}$ \\ \hline
$II:\ 0.5398\leq d < 0.6145$  & $2\omega_g,\ \omega_g+ \omega_s $ & 
       $P_t=-\ve^2P\{P,Q\} ,$ \\ 
$2\omega_g\in\sigma_{cont}$  & $2 \omega_s-\omega_g$ & $Q_t=-\ve^2PQ\{1,
\ve^2\{Q\}\}$ \\ \hline
$III:\ 0.6145\leq d < 0.6364$  & $2\omega_g,\ 2 \omega_s-\omega_g $ & 
       $P_t=-\ve^2\{P^2\},$ \\ 
  &  & $Q_t=-\ve^4\{PQ^2\}$ \\ \hline
$IV:\ 0.6364 \leq d<0.6679$ & $\omega_g+ \omega_s$ & $P_t=-\ve^2\{PQ\}$, 
  \\ $2\omega_g\notin\sigma_{cont}$ & 
 $2 \omega_s-\omega_g$ &       $Q_t=-\ve^2\{PQ
\{1,\ve^2\{Q\}\}\}$\\ \hline
$V:\ 0.6679 \leq d< d_e$ & $\{3\omega_g, 4\omega_g,2\omega_s-\omega_g\}$ &
          $P_t=-\ve^4P^3\{1,\ve^2P\}-\ve^2P Q\{ 1,\ve^2P\},$ \\ 
  $d_e\sim 0.82$  &     $\{\omega_g+\omega_s,\ 2\omega_g+\omega_s\}$ &
    $Q_t=-\ve^2P Q\{ 1,\ve^2\{P,Q\}\}$\\ \hline 
$VI:\ d_e \leq d <0.9229 $ & $ \{4\omega_g, 5\omega_g ,2\omega_s-\omega_g\}$
& $P_t=-\ve^2P\{Q\{1,\ve^2P\},R\{1,\ve^2P\},\ve^4 P^3\}+...$ \\
  $\omega_e$ appears & 
  $\omega_g + \omega_s, 2\omega_g+\omega_s,2\omega_e-\omega_s$ &
  $Q_t= -\ve^2PQ\{1,\ve^2\{P,Q,R\}\} +...$\\
     & $\omega_g+\omega_e, 2\omega_g+\omega_e,\omega_e+\omega_s-\omega_g$ & 
 $R_t=\ve^2PR\{ 1, \ve^2\{P,Q\}\} -\ve^4QR^2 +...$ \\ \hline
$VII:\ 0.9229\leq d< d_*$ & $ \{n\omega_g\}_{5\le n\le 38}, 2\omega_s$&
$P_t= -\ve^{2n-2}\{P^n\} +...$ \\
$d_*\sim 1.2234$ & $\omega_g+\omega_s, \omega_g+\omega_e, 2\omega_g+\omega_s,$& 
$Q_t= -\{\ve^2Q^2\}+...$\\
$2\omega_s\in\sigma_{cont}$ 
 & $2\omega_g+\omega_e, 2\omega_g+2\omega_s, \omega_g+2\omega_s,$&
$R_t= -\ve^2R\{Q,P\}+...$\\
& $\omega_g+\omega_s+\omega_e, \omega_s+\omega_e,2\omega_e-\omega_s$ &    \\ 
& $2 \omega_e-\omega_g,\omega_e+\omega_s-\omega_g,2\omega_s-\omega_g$ & 
\\ \hline
$VIII:\ d_*\le d $ & $n\omega_g\ (5\le n\le N), 2\omega_s, 2\omega_e$ &
$P_t= -\ve^{2n-2}\{P^n\}+...$ \\
$2\omega_e\in\sigma_{cont}$ &  & $Q_t= -\{\ve^2Q^2\}+...$\\
& & $R_t=-\{\ve^2R^2\}+...$ \\ \hline
\end{tabular}
\end{center}
}

\noindent{\bf Remark on notation:} We illustrate the notation of the 
table with the following example. Consider the regime V. This 
regime can be broken into three subregimes illustrated in the following
table: \\

\bigskip
{\small
\centerline{\bf Table 3: internal mode power equations for subregime $V$ }
}
{\footnotesize
\begin{tabular}{|l|l|l|} \hline
Subregime of $d$ & Resonances &  System Form \\ \hline
$a)\ 0.6679 \leq d <0.743$ & $3 \omega_g, \omega_g+\omega_s$ &
$P_t=-\ve^2C_1 P Q -\ve^4C_2 P^3,$ \\
& $2 \omega_s-\omega_g$ & $Q_t=-\ve^2C_3 P Q -\ve^4C_4 P Q^2$ \\ \hline
$b)\ 0.743 \leq d <0.7622$ & $3 \omega_g, 4\omega_g$ &
$P_t=-\ve^2C_1 P Q -\ve^4C_2 P^3-\ve^6 C_5 P^4,$\\
    & $\omega_g+\omega_s, 2\omega_s-\omega_g$ &  $Q_t=-\ve^2C_3 P Q
-\ve^4 C_4 P Q^2 $ \\ \hline
$c)\ 0.7622 \leq d <0.7687 $ & $ 3 \omega_g, \omega_g + \omega_s, 2 \omega_s -
\omega_g $
& $P_t=-\ve^2C_1 P Q -\ve^4C_2 P^3- \ve^4C_6P^2 Q-\ve^6C_5P^4$ \\
     & $4 \omega_g,2 \omega_g+\omega_s$ & $Q_t=-\ve^2C_3 P Q- \ve^4C_4 P Q^2-\ve^4C_7 P^2 Q $ \\ \hline
$c)\ 0.7687 \leq d<d_e$ & $ \omega_g+\omega_s,2\omega_g+\omega_s$ &
$P_t=-\ve^2C_1 P Q -\ve^4C_6P^2 Q-\ve^6C_5P^4$ \\
 $d_e\sim0.82$     & $4 \omega_g, 2 \omega_s-\omega_g$    & $Q_t=-\ve^2C_3 P Q -\ve^4C_4 PQ^2 -\ve^4C_7 P^2 Q$  \\ \hline
\end{tabular}
}
\bigskip

\noindent
 Although the details of the system form change between regimes,
the topological character of the phase portrait and therefore the 
qualitative nature of the solutions does not change; 
 each phase portrait
has $P=0$ as a stable line of equilibria. 
For regime III, Table 2 is read as follows: in this regime some or 
all resonances occur from among each of the indicated sets:
$\{3\omega_g,4\omega_g\}$, and 
 $\{\omega_g+\omega_s, 2\omega_s-\omega_g, 2\omega_g+\omega_s\}$, giving rise to terms
 in the following manner:
\be k\omega_g+l\omega_s+m\omega_e \ \rightarrow\ \ve^{2n-2} P^kQ^lR^m, 
\ \  k,l,m\in\Z_+, \ \ k+l+m=n.\ee
Thus the sets in curly brackets in Table 2
 are to be viewed as columns of a ``menu"
from which one (or the dynamical system) chooses all or some items
 (resonant combinations) depending on the subregime of $d$. 
 For a given subregime, this choice of subset 
 gives rise to linear combination with {\it nonnegative coefficients}, 
 $C_j$,
 of the corresponding monomials in
 $P,Q$ and $R$ in the power equations. Therefore 
 $P_t,Q_t,R_t\le0$.  
 Generically , we have $C_j>0$. 
\medskip

It is straightforward to analyze the sets of equilibria and their
  stability for each of the 
systems in Table 2. These take the form of constant vectors with 
{\it at most} one nonzero component. These states are dynamically 
 stable. 
 If we take the view that the system of power 
 equations  determines  the nonlinear dynamics near the static kink
we anticipate that:

\begin{itemize}
\item the zero solution of the internal mode power equations
corresponds to the ground state
  static kink solution discrete nonlinear equation. 
We denote the ground state kink by $K_{gs}$. 

\item a nonzero equilibrium state with $Q\ne0$   
corresponds to a time periodic solution:
\be u_i \sim K_{gs,i} + \cos({\omega_s}t) s_i.\nn\ee
We call such a periodic solution of the full nonlinear dynamical system,
which would have the same symmetry as the kink, a {\it wobbling kink},
 which we designate by $sW$ or simply $W$.

\item a nonzero equilibrium state with $P\ne0$  power equations
corresponds to a time periodic solution:
\be u_i \sim K_{gs,i} + \cos({\omega_g}t) g_i.\nn\ee
 We call such a periodic 
solution a {\it g-wobbling kink}. We denote this state by $gW$.

\item a nonzero equilibrium state with $R\ne0$  power equations
corresponds to a time periodic solution:
\be u_i \sim K_{gs,i} + \cos({\omega_e}t) g_i.\nn\ee
Such a periodic solution of the full nonlinear dynamical system, 
would not have the same symmetry as the kink. We call such a periodic 
solution an {\it e-wobbling kink}. We denote this state by $eW$.
\end{itemize}

Below, we present a table of the kinds of static and periodic states 
 anticipated
by the normal form / internal mode power equation analysis.
\bigskip

{\small
\noindent{\bf Table 4: $\phi^4$ normal form, equilibria and anticipated
coherent structures}
}
{\footnotesize
\begin{center}
\begin{tabular}{|l|l|l|} \hline
Regime of $d$& Equilibria & Coherent structures \\ \hline
I:\ $ d < 0.5398$  &$\{(P,0):P\ge0\}$  &$K_{gs},\ W, gW$\\
                          &$\{(0,Q):Q\ge0\}$  &
  \\ \hline
II-III:\ $0.5398\leq d < 0.6364$  &$\{(0,Q):Q\ge0\}$  &$K_{gs},\ W$ 
  \\ \hline
IV:\ $0.6364 \leq d<0.6679$ &$\{(0,Q):Q\ge0\}$  &$K_{gs},\ W,\ gW$\\
                            &$\{(P,0):P\ge0\}$ &
  \\ \hline
V:\ $0.6679 \leq d <d_e$ &  $\{(0,Q):Q\ge0\}$  & $K_{gs},\ W$\\
 $d_e\sim0.82$ & & \\ \hline
VI:\ $d_e \leq d <0.9229 $ & $\{(0,Q,0):Q\ge0\}$  &$K_{gs},\ W,\ eW$\\ 
			& $\{(0,0,R):R\ge0\}$ &  \\ \hline
VII:\ $0.9229\leq d<1.2234$ & $\{(0,0,R):R\ge0\}$ &$K_{gs},\ eW$ \\
$d_*\sim 1.2234$ & & \\ \hline
VIII:\ $d_*\le d $ & $\{(0,0,0)\}$ &$K_{gs}$\\ \hline
\end{tabular}
\end{center}
}
\bigskip

\subsection{The normal form for discrete sine-Gordon}

The procedure for deriving the normal form for the internal mode amplitudes 
and the internal mode power equations presented in sections 3 and 4.1  
 can be applied to the discrete sine-Gordon
equation as well. The implementation is actually simpler because for 
 discrete SG there are only two internal modes: $g$ and $e$; see section 2.
 Therefore, the decomposition of the solution is:
\be
u_i(t) = K_i + \ve a(t)g_i + \ve b(t) e_i + \ve^2\eta_i(t),\label{eq3SG}\ee
where 
\ba &&
\lan g,\eta(t)\ran=\lan e,\eta(t)\ran = 0\nn\\
&&
 P_c\eta\equiv \eta -\lan g,\eta(t)\ran g -\lan e,\eta(t)\ran e\ =\ \eta(t)
\nn\ea

As in the previous subsection the amplitude equations for the slowly varying 
modulation functions can be obtained:

\ba
A_t &=& \ve^2\left(\alpha_1|A|^2 + \alpha_2|B|^2\right) A +
  \ve^4\alpha_3 |B|^4A + \Xi_A(A,B,t)\nn\\
B_t &=& \ve^2\left(\beta_1|A|^2 + \beta_2|B|^2\right) A +
  \ve^4\beta_3 |B|^4A + \Xi_B(A,B,t)\label{eq:dSGAB}
\ea
The coefficients $\alpha_j, \beta_j$ can be evaluated along the lines
 detailed in the pervious section and are tabulated below.

\bigskip

{\small
\centerline{\bf Table 5A : Principal SG g-mode coefficients}
}
{\footnotesize
\begin{center}
\begin{tabular}{|c|c|} \hline
$|A|^2 A$ &  $-{1\over8}d^{-4}{\omega_g}^{-1}\lan \sin{K}g^2,
G(2\omega_g) \sin{K}g^2\ran$ \\ \hline
$|A|^4 A$ &  $-{3\over 4(3!)^2} d^{-4}\omega_g^{-1}\lan \cos{K}g^3, 
 G(3\omega_g) \cos{K}g^3\ran$ \\
\hline
$|B|^2A$      & $-{1\over2}d^{-4}\omega_g^{-1}\lan \sin{K}ge,
 G(\omega_g+\omega_e) \sin{K}ge\ran$\\ \hline
$|B|^4A$      & $-{9\over 4(3!)^2}d^{-4}{\omega_g}^{-1}\lan \cos{K}ge^2, 
 G(\omega_g+2\omega_e) \cos{K}ge^2\ran $\\ \hline
$|A|^2|B|^2A$ & $-{18\over 4(3!)^2} d^{-4}\omega_g^{-1} \lan\cos{K}g^2e,
 G(2\omega_g+\omega_e)\cos{K}g^2e>$\\ \hline
\end{tabular}
\end{center}
}
\bigskip

{\small
\centerline{\bf Table 5B : Discrete SG e-mode coefficients}
}
{\footnotesize
\begin{center}
\begin{tabular}{|c|c|} \hline
$|B|^2 B$ &  $-{1\over8}d^{-4}\omega_e^{-1}\lan
  \sin{K}e^2,G(\omega_e)
 \sin{K}e^2\ran $ \\ \hline
$|B|^4 B$ &  $-{3\over 4(3!)^2}d^{-4}\omega_e^{-1} \lan \cos{K}e^3,
 G(3\omega_e) \cos{K}e^3\ran $ \\ \hline
$|A|^2B$      & $-{1\over2}d^{-4}\omega_e^{-1} \lan \sin{K}ge,
G(\omega_g + \omega_e)
 \sin{K}ge\ran$\\ \hline
$|A|^4B$      & $-{9\over 4(3!)^2} d^{-4} \omega_e^{-1} 
 \lan \cos{K}eg^2,
G(\omega_e+2\omega_g) \cos{K}eg^2\ran$\\ \hline
$|A|^2|B|^2B$ & $-{18\over 4 (3!)^2} d^{-4}\omega_e^{-1} \lan\cos{K}e^2g,
G(\omega_g+2\omega_e)\cos{K}e^2g\ran $\\ \hline
$|A|^2|B|^2B$ & $-{18\over 4 (3!)^2} d^{-4}\omega_e^{-1} \lan\cos{K}e^2g,
G(2\omega_e-\omega_g)\cos{K}e^2g\ran $\\ \hline
\end{tabular}
\end{center}
}
\bigskip

As with discrete $\phi^4$, the details of the internal mode power equations
change as $d$ varies due to the different types of resonances with the 
continuous spectrum which may occur. 
 For discrete SG there are essentially only two 
regimes, and one value of the parameter $d$ across which there is a topological
change in the phase portraits. Since the detailed picture is simpler
we tabulate it in greater detail.

 Figure 7  displays the variation of the internal mode 
frequencies $(\omega_g, \omega_e)$, certain multiples of them 
  and certain other integer
linear combinations of them. As with discrete $\phi^4$, transitions in 
 the structure of the normal form occur across values of $d$ 
 for which there is a change in the set of integer linear combinations which 
 lie in the band of continuous spectrum. 
\begin{figure}[t]
\centerline{\bf Arithmetic of SG kink frequencies}
\centerline{\psfig{file=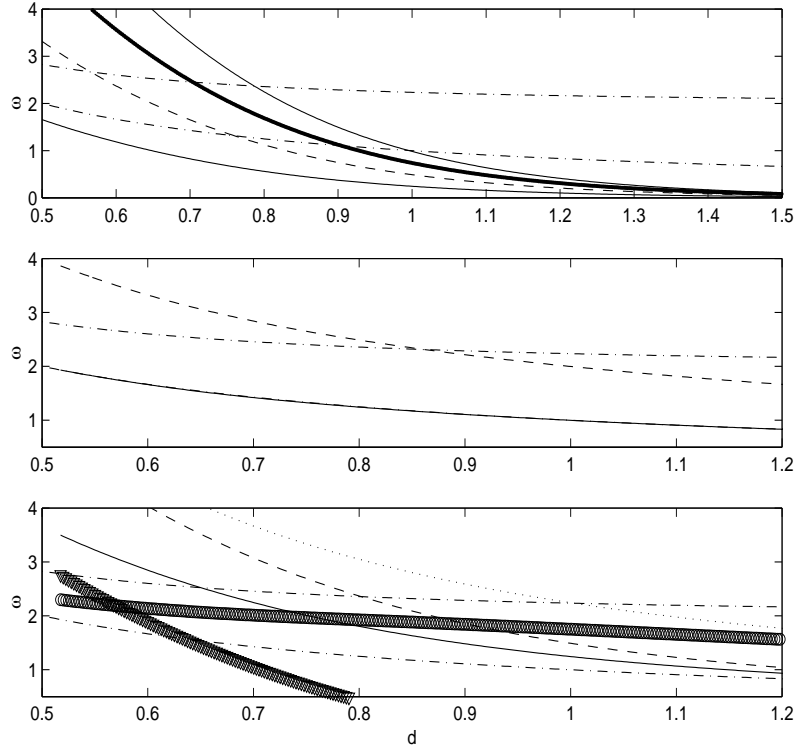,height=4.0in,width=4.2in,angle=0}}
\caption{Top panel shows the Goldstone harmonics (solid:
first, dashed: second, thick:third and again solid: fourth).
Higher harmonics
of $g$ will also be relevant, but are not shown here.
Middle panel shows the shape mode (solid line) that  lives
close to the band edge (maximum bifurcation is of the order of $0.01$)
and its second harmonic (dashed). Finally, lower panel shows
mixed combinations of  harmonics  (solid:
to $\omega_g+\omega_s$, dashed: $2 \omega_g+\omega_e$ and
dotted to $\omega_g+2 \omega_e$). The circles denote 
$2 \omega_e-\omega_g$ while the down triangles $3\omega_g-\omega_e$
-see note below-. In all cases the phonon band edges are shown 
by dash-dot lines.}
\label{fig:SGcombos}
\end{figure}
The precise normal form and 
 power equations can be worked
out using the coefficient Table 5 and the expression for 
 $G(\zeta)$, (\ref{eq:Gzeta}). Notice that only the leading order
terms are given in these internal mode power equations.

{\footnotesize
\centerline{\bf Table 6: SG internal mode power equations}
\begin{center}
\begin{tabular}{|l|l|l|} \hline
Regime of $d$& Resonances &  System Form \\ \hline
I:\  $d<d_e\sim 0.515$ &  None & $P_t=0$\\ \hline
 II:\  $d_e\le d<.565$ & $2\omega_e-\omega_g,\{3\omega_g-\omega_e\}$& 
$P_t=-\ve^6 \{P^3Q\},\ 
Q_t = -\ve^4\{PQ^2\}$\\ \hline
 III:\ $0.565\leq d <0.65$& $2 \omega_g,2\omega_e-\omega_g$ & 
 $P_t=-\ve^2\{ P^2\}$, $Q_t=- \ve^4\{PQ^2\}$ \\ \hline
 $ 0.65\leq d<0.7$& $2 \omega_g,\omega_g+\omega_e$ & 
 $P_t=-\ve^2 P \{P, Q\}$,\\ 
 &    &               $Q_t=-\ve^2 P Q\{ 1,\ve^2Q\}$
\\ \hline
 $0.7\leq d < 0.76$& $2 \omega_g, 3 \omega_g, \omega_g+\omega_e,$ 
& $P_t=-\ve^2 P\{ P, Q, \ve^4P^2\}$, \\
 & $2\omega_e-\omega_g$          &  $Q_t=-\ve^2PQ\{ 1, \ve^2Q\}$ \\ \hline
 $0.76\leq d <0.785$& $3 \omega_g, \omega_g+\omega_e,$ 
& $P_t=-\ve^2P \{ Q\} - \ve^2\{P^2\}$,  \\
         &   $2\omega_e-\omega_g$     & $Q_t=-\ve^2PQ\{1,\ve^2Q\}$ 
\\ \hline
 $0.785\leq d <0.8$&  $3 \omega_g, 4 \omega_g$ & $P_t=
 -\ve^2P\{Q,\ve^2P^2,\ve^6P^4\}$, \\
       & $\omega_g+ \omega_e,2\omega_e-\omega_g$   &               
$Q_t=-\ve^2PQ\{1, \ve^2P Q\}$ \\ \hline
 $0.8\leq d < 0.847$& $3 \omega_g, 4 \omega_g,2\omega_e-\omega_g$
& $P_t=-\ve^2P\{ Q, \ve^2 P^2, \ve^2 PQ, \ve^4 P^3\}$, \\
    &  $\omega_g+\omega_e,2\omega_g+\omega_e$       &          
$Q_t=-\ve^2PQ\{1, \ve^2Q, \ve^2P\}$ \\ \hline
 $0.847\leq d <d_*$ & $3 \omega_g,4 \omega_g,5 \omega_g,2\omega_e-\omega_g$ 
& $P_t=-\ve^2P\{Q, \ve^2P^2,\ve^4 P^3, \ve^2P Q, \ve^6 P^4\}$,\\
 $d_*\sim 0.86$     &  $\omega_g+ \omega_e,2 \omega_g+ \omega_e$  &  
$ Q_t=-\ve^2PQ\{ 1, \ve^2 Q,\ve^2P\}$ \\ \hline
 IV:\ $d_*\leq d < 0.9$& $3 \omega_g, 4 \omega_g,5 \omega_g, 2\omega_e$ & 
 $P_t=-\ve^2P\{ Q, \ve^2P^2, \ve^4P^3, \ve^2PQ\}$, \\
    & $\omega_g+\omega_e,2 \omega_g+\omega_e,2\omega_e-\omega_g$         
&      $Q_t=-\ve^2Q\{ P, Q, \ve^2 PQ, \ve^2P^2\}$ \\ \hline
$0.9\leq d<0.99$&  $4 \omega_g, 5 \omega_g, 2 \omega_e$ & 
 $P_t=-\ve^2P\{ Q, \ve^4P^3, \ve^2PQ, \ve^4P^4\}$,  \\
    & $\omega_g+\omega_e, 2 \omega_g + \omega_e,2\omega_e-\omega_g$        &
  $Q_t=-\ve^2Q\{ P, Q, \ve^2PQ, \ve^2P^2\}$ \\ \hline
$0.99\leq d <1.003$& $5 \omega_g, 2 \omega_e,2\omega_e-\omega_g$ & 
 $P_t=-\ve^2P\{Q, \ve^2PQ, \ve^6P^4\}$,  \\
  & $\omega_g+ \omega_e, 2 \omega_g+ \omega_e$    &    
$Q_t=-\ve^2Q\{ Q, \ve^2PQ, \ve^2P^2\}$ \\ \hline
$1.003\leq d $& $ n\omega_g\ (n\ge5), 2 \omega_e,\omega_g+\omega_e$ & 
 $P_t=-\ve^2P\{Q,\ve^2PQ,\ve^{2n-2}P^{n-1},\ve^2Q^2,...\}$, \\
  & $2 \omega_g+\omega_e,\omega_g+2 \omega_e$   & 
  $Q_t=-\ve^2Q\{P,Q, \ve^4P^2, \ve^4PQ,...\}$ \\ \hline 
\end{tabular}
\end{center}
}

The inferred coherent structures are displayed in the following 
table.
\bigskip

{\small
{\bf Table 7:\  SG normal form, equilibria and anticipated coherent structures}
}
{\footnotesize
\begin{center}
\begin{tabular}{|l|l|l|} \hline
Regime of $d$& Equilibria & Coherent structures\\ \hline
I:\ $d<d_e$ & $P\ge0$ & $K_{gs},\ gW$\\ \hline
II:\ $d_e\le d <0.565$& $\{(P,0),(0,Q): P\ge0,\ Q\ge0\}$& $K_{gs},\ W,\ 
gW$\\ \hline
III:\ $0.565\leq d < d_*$  & $\{(0,Q):Q\ge0\}$& $K_{gs},\ W$
  \\ 
 $d_*\sim 0.86$ & &\\ \hline
IV:\ $d_* \leq d$ & $\{ (0,0)\}$ &$K_{gs}$
  \\ \hline
\end{tabular}
\end{center}
}
\bigskip

\section{ Time periodic solutions}
\setcounter{equation}{0}

\subsection{Existence of wobbling kinks}

From the discussion of the previous section, 
 we expect that the dispersive 
 normal form we have derived for discrete 
 nonlinear wave equations  (in particular, discrete SG and $\phi^4$)
models the dynamics in a neighborhood of the static kink solution.
 This normal form captures the effects of nonlinear 
resonant interactions of the internal and continuum mode fluctuations
about the kink. Tables 4 and 7 indicate for  discrete $\phi^4$  discrete SG
 the anticipated existence of periodic solutions which bifurcate from the 
static kink solution. In this section we prove, under suitable nonresonance
hypotheses, the existence of such time periodic solutions.

In tables 3 and 5 we see that these normal 
 forms fall into two categories (i) those for which the origin is the only 
 equilibrium ($d$ sufficiently large, $d>d_*$) 
 (ii) $d$ is such that the normal form
 admits one or two lines of equilibria. In case (i) we see that the  
 zero solution is asymptotically stable. By (\ref{eq3}) and (\ref{eq3SG})
  this suggests that the ground state
kink is a stable attractor.
  At $d_*$ there is 
 a bifurcation in the phase portrait; for $d<d_*$, zero is 
 no longer an isolated equilibrium. Rather, the normal form now has a  
line of stable equilibria passing through the origin. This then suggests  
an approximate solution of the nonlinear wave equation of the form:
\be u_j \sim K_j\ +\ 
\ve B_{eq}\cos(\Omega t)\xi_{\Omega,j},  
\ \  j\in\Z
\label{eq:approx}\ee
where $\xi_\Omega$ denotes an internal mode  $\Omega$ its corresponding  
 frequency; $B\chi_\Omega = \Omega\chi_\Omega$ and $\ve$ is small. 
If such a family of periodic solutions exists, we say this family bifurcates
from the kink solution in the direction of the internal mode, $\xi_\Omega$. 
Such directions for bifurcation are available for $d<d_*$.

\noindent{\it For $d<d_*$ are there bifurcating time 
 periodic solutions?}
 
In view of the  analysis of the regime $d>d_*$,
  one must be cautious about reaching a conclusion about the large
time behavior on the basis of the above approximation. In fact we learn from
the regime $d>d_*$ that radiation damping, due to resonant coupling of
oscillations to the continuum, may lead  
  to the slow 
decay of a solution which appears to be time-periodic on shorter 
time scales. 

On the other hand, given the approximate periodic solution (\ref{eq:approx})
  it is certainly natural to 
attempt a perturbative (in $\epsilon$) 
 construction of a true time-periodic solution.
 The approach we use is the Poincar\'e continuation, which is standard 
in the persistence theory of periodic solutions of systems of 
 ordinary differential equations \cite{CL}.  
 As in typical perturbation expansions, one expects a hierarchy of 
linear inhomogeneous problems governing contributions to the expansion 
at various orders in $\epsilon$. Solvability at each order requires
 that one can arrange for the solution to each inhomogeneous problem to
be time periodic  of some common period.
 That this can be 
achieved in the problem at hand follows from the following 
 nonresonance condition, 
 verified in our numerical investigation of the spectrum of the kink: 

\noindent{\bf (NR)}
 {\it Assume that no integer multiple of the internal mode frequency $\Omega$
lies in the continuous spectrum of $B$.} 
  
A precise statement of the result is as follows:

\begin{theo}
Let $ K=\{K_i\}_{i\in\Z}$ denote a ground state kink. Assume that the 
 linear operator, $B$, 
acting on the space $l^2(\Z)$, 
has a simple eigenvalue, $\Omega$, with corresponding 
 eigenfunction, $\xi_\Omega$ satisfying $B\xi_\Omega=\Omega\xi_\Omega$.
Also assume the nonresonance condition {\bf (NR)}. Then, in a neighborhood of 
 $K$, there is a curve
of solutions $\epsilon\mapsto y(t;\epsilon)$ passing through the ground 
state kink with the following properties:
\begin{itemize}
\item There is a number $\epsilon_0>0$ such that 
 $y(t;\epsilon)$ is defined for all $\epsilon<\epsilon_0$.
\item $y(t;0)=K$
\item $y(t;\epsilon)$ has period $2\pi/\omega(\epsilon)$ in $t$, 
where
  $\omega^2(\epsilon)=\Omega^2(1+{\cal O}(\epsilon))$ 
 is a smooth function with $\omega(0)=\Omega$.
\item $y(t;\epsilon)\in l^2(\Z)$
\item $ y_i(t;\epsilon) = K_i + \epsilon\cos({\omega(\epsilon) t})
 \xi_{\Omega,i} + {\cal O}(\epsilon^2).\label{eq:expansion}$
\end{itemize}
\end{theo} 

The next result, a simple consequence of the proof of Theorem 5.1,
yields the class of solutions which we call {\it wobbling kinks}.
\begin{cor}
For discrete SG, the periodic solution bifurcating from the kink
in the direction of the spatially odd
 edge mode ($Be=\omega_ee$) 
has the same symmetry as the  SG kink around its center.
 For discrete $\phi^4$ the periodic solution bifurcating from the kink
in the direction of the spatially odd shape mode ($B s=\omega_s s$)
has the same symmetry as the $\phi^4$ kink.
\end{cor}

We seek a solution in the form:
\be
u_i=K_i+ b(t) \xi_{\Omega,i} + \eta_i,\ i\in\Z.
\label{eq37}
\ee
Substitution of (\ref{eq37}) into (\ref{dphi4}) and then projection of 
the resulting equation separately onto the eigenvector $\xi_\Omega$
  and its orthogonal
complement yields the coupled system for the shape mode amplitude and the 
radiation components
\footnote{Here we employ the Taylor expansion for the function $W=V'$:
\be W(K+M) = W(K) + W'(K)M + \int_0^1(1-\theta)W''(K+\theta M) d\theta M^2
\nn\ee}
\ba
\D_t^2b+ \Omega^2 b&=&-d^{-2}
 \lan \xi_\Omega,{\cal V}[b\xi_\Omega+\eta](b\xi_\Omega+\eta)^2\ran
\label{eq38}\\
\D_t^2\eta+ B^2 \eta&=&-d^{-2}P_c{\cal V}[b\xi_\Omega+\eta](b\xi_\Omega+\eta)^2,
\label{eq39}
\ea
where
\be
{\cal V}[M]\ =\ \int_0^1 (1-\theta)V'''(K+\theta M) d\theta.
 \label{eq:calVdef}\ee

\begin{rmk}
It is possible to formulate the proof of Theorem 5.1, as essentially 
 a consequence results in \cite{CL} (Chapter 14, Theorem 2.1),  
 applied 
to (\ref{eq38}-\ref{eq39}); a simple generalization of the result 
to an infinite dimensional setting is required.
  In this context,  hypothesis {\bf (NR)},  
corresponds to the hypothesis that  $1$ be a simple Floquet multiplier
of the unperturbed linear problem obtained by setting the right hand sides
of (\ref{eq38}-\ref{eq39}) equal to zero. In this paper we 
 present a direct proof  of Theorem 5.1 
 in order to make our study elementary and self-contained. 
\end{rmk} 
We shall be considering {\it small} perturbations of the static kink, and 
 therefore 
 introduce rescaled internal mode 
and radiation components: 
\be b=\epsilon b_1,\ \ \eta= \epsilon \eta_1,\nn\ee 
where $\epsilon$ is a small parameter.
Since the nonlinear wave equations considered are autonomous, we can't
 {\it a priori} specify the period, we leave the 
period unspecified at this stage and introduce a new time variable, $\tau$,
with respect to which the sought-for periodic solution is $2\pi$ periodic.
Let
\be
t=\omega(\epsilon)^{-1}\tau,\ \omega(0)=\Omega,
\label{eq40}
\ee
and 
\be b_1(t)=\beta_1(\tau),\ \beta_1(\tau+ 2 \pi) = \beta_1(\tau).\nn\ee

Then, equations  
(\ref{eq38}-\ref{eq39}) become:
\ba
\left(\omega^2(\epsilon)\D_\tau^2+ \Omega^2\right)
\beta_1(\tau)&=&\epsilon {\cal F}_{\beta}(\beta_1,\eta_1; \epsilon)\nn\\
{\cal F}_{\beta}(\beta_1,\eta_1; \epsilon)&\equiv& 
-d^{-2}\lan \xi_\Omega, {\cal V}[\epsilon(\beta_1\xi_\Omega+\eta_1)] 
(\beta_1\xi_\Omega+\eta_1)^2\ran\nn\\
\label{eq41}\\
\left(\omega^2(\epsilon)\D_\tau^2+ B^2\right) \eta_1(\tau,i)&=&
\epsilon {\cal F}_\eta (\beta_1,\eta_1;\epsilon)\nn\\
{\cal F}_\eta(\beta_1,\eta_1;\epsilon) 
 &\equiv&-d^{-2}P_c {\cal V}[\epsilon(\beta_1\xi_\Omega+\eta_1)] 
 (\beta_1\xi_\Omega+\eta_1)^2 
\label{eq42}
\end{eqnarray}

Let
\begin{eqnarray}
\omega^2(\epsilon)=\Omega^2 (1+ \epsilon \sigma(\epsilon)),
\label{eq43}
\end{eqnarray}
where $\sigma(\epsilon)$ is to be determined. Then (\ref{eq41}) becomes
\begin{eqnarray}
\left(\D_\tau^2+1\right)\beta_1=\frac{\epsilon}{1+\epsilon \sigma(
\epsilon)}[\sigma(\epsilon) \beta_1 + \Omega^{-2}
{\cal F}_{\beta}(\beta_1,\eta_1;\epsilon)].
\label{eq46}
\end{eqnarray}
We extract the leading order behavior by defining:
\be \beta_1=\cos(\tau)+ \epsilon \beta_2,\nn\ee
where
\be
({\partial_{\tau}}^2+1) \beta_2=\frac{1}{1+\epsilon \sigma(\epsilon)}
[\sigma(\epsilon) \cos(\tau)+ \epsilon \sigma(\epsilon) \beta_2
+\Omega^{-2} {\cal F}_{\beta}\left(\cos(\tau) + \epsilon
\beta_2,\eta_1;\epsilon\right)
\label{eq47}
\ee
By the Fredholm alternative, equation (\ref{eq47}) has an even $2\pi$ periodic
 solution  if and only if 
\be
P_{\cos}(\sigma \cos(\tau)+ \epsilon \sigma \beta_2+
\Omega^{-2} {\cal F}_{\beta})=0.
\label{eq48}
\ee
Here, 
\begin{eqnarray}
P_{\cos}\ g = \pi^{-1}\int_0^{2 \pi} \cos(\mu) g(\mu) d\mu
\label{eq49}
\end{eqnarray}
is the projection onto the even $2\pi$ periodic null space of $\D_\tau^2+1$.
The orthogonality constraint (\ref{eq48}) can be rewritten as 
\be
\pi \sigma + \epsilon \sigma
\int_0^{2 \pi} \cos(\tau) \beta_2(\tau) d \tau
+\omega_s^{-2} \int_0^{2\pi} \cos(\tau) {\cal F}_\beta
\left(\cos(\tau)+ \epsilon \beta_2(\tau), \eta_1;\epsilon\right) d\tau=0
\label{eq50}
\ee
We have therefore reformulated the problem of finding a 
periodic solution of the discrete nonlinear wave equation, or equivalently
(\ref{eq38}-\ref{eq39}), as the problem of finding $2\pi$ periodic solutions
in $\tau$, 
 $(\beta_2(\epsilon),\eta_1(\epsilon),\sigma(\epsilon))$, of the system:
 \be F(\beta_2,\eta_1,\sigma;\epsilon)=0,\nn\ee
where $F= \left( F_1,F_2,F_3\right)^{t}$, is defined by:
\ba
&&F_1(\beta_2,\eta_1,\sigma;\epsilon)\ =\ 
(\D_\tau^2+1)\beta_2\nn\\
&&-(1+\epsilon \sigma)^{-1}
\left[\sigma \cos(\tau)+ \epsilon \sigma\beta_2
+\Omega^{-2} {\cal F}_\beta\left(\cos(\tau) + \epsilon
\beta_2,\eta_1;\epsilon\right)\right]\label{eq:F1}\\
&& F_2(\beta_2,\eta_1,\sigma;\epsilon)\ =\ 
\left(\Omega^2(1+\epsilon\sigma)\D_\tau^2+ B^2\right) \eta_1(\tau,i)
-\epsilon {\cal
  F}_\eta (\cos(\tau) + \epsilon\beta_2,\eta_1;\epsilon)\nn\\
&&\label{eq:F2}\\
&& F_3(\beta_2,\eta_1,\sigma;\epsilon)\ =\ 
-(1+\epsilon\sigma)^{-1}\left[\pi \sigma + \epsilon \sigma
\int_0^{2\pi} \cos(\tau) \beta_2 d \tau\right.\nn\\
&&\ \ \ +\left. \Omega^{-2} \int_0^{2 \pi} \cos(\tau) {\cal F}_{\beta}
(\cos(\tau)+ \epsilon \beta_2(\tau), \eta_1;\epsilon)d \tau\right]\label{eq:F3}
\ea

We view $F$ as a mapping of $(\zeta,\epsilon)\in{\cal X}
\mapsto F(\zeta,\epsilon)\in{\cal Y}$. 
Here, ${\cal X}$ and ${\cal Y}$ are defined by: 
\ba {\cal X} &:&\  \zeta=(\beta,\eta,\sigma)\ {\rm such\ that}\nn\\   
 && \beta=\beta(\tau)\in H^2,\ even,\ {\rm and}\ 2\pi\ {\rm periodic}\nn\\
 && \eta=\eta(\tau,\cdot)\in H^2 \ even,\ {\rm and}\ 2\pi\ {\rm periodic
 \ in}\ \tau\nn\\
&&  {\rm with\ values\ in
\ the\  space\  of\  }\ l^2(\Z)\ 
 {\rm functions\ and}\  
\sigma\in\R\nn\\
{\cal Y}&:&  \zeta=(\beta,\eta,\rho)\ {\rm such\ that}\nn\\
&& \beta=\beta(\tau)\in L^2,\ even,\ {\rm and}\ 2\pi\ {\rm periodic}\nn\\
&& \eta=\eta(\tau,\cdot)\in L^2 \ even,\ {\rm and}\ 2\pi\ {\rm periodic
 \ in}\ \tau\nn\\
&& {\rm with\ values\ in
\ the\  space\  of\ }\ l^2(\Z)\  {\rm functions\ and}\nn\\ 
&& \rho=\int_0^{2\pi} \beta(\mu)\cos(\mu) d\mu.\nn\ea

We find a particular $\zeta^{(0)}\in{\cal X}$ for which $ F(\zeta^{(0)};0)=0$,
and then seek to construct curve of solutions $\epsilon\mapsto\zeta(\epsilon),
 \ \zeta(0)=\zeta^{(0)}$, for all sufficiently small 
 $\epsilon$ using the implicit
function theorem \cite{Nirenberg}. 
  To find $\zeta^{(0)}$ we set $\epsilon=0$ and consider
 the system
$F(\zeta;0)=0$. Taking $\eta_1=\eta_1^{(0)}\equiv0$ and $\sigma=\sigma^{(0)}=0$, 
 we find that $\beta_2^{(0)}$ satisfies the 
equation:
\be \left(\D_\tau^2+1\right)\beta_2^{(0)}=-d^{-2}\Omega^{-2}
 \lan \xi_\Omega,{\cal V}[0] \xi_\Omega^2\ran\cos^2(\tau),\nn\ee
which has the solution
\be \beta_2^{(0)} =
  -{1\over2}d^{-2}\Omega^{-2}\lan \xi_\Omega\,{\cal V}[0]\xi_\Omega^2\ran 
 \left(1-{1\over3}\cos(2\tau)\right).\nn\ee
Thus, $\zeta^{(0)}=(\beta_2^{(0)},0,0)\in{\cal X}$ satisfies
 $F(\zeta^{(0)};0)=0$. We shall now  use the implicit 
function theorem
to continue this solution to show that this solution deforms uniquely to 
nearby solutions for $\epsilon$ sufficiently small and nonzero.

To apply the implicit function theorem, it suffices to check that 
 $d_{\zeta}F
(\zeta^{(0)};0)$ is bounded and invertible. A computation yields: 
\be
d_{\zeta} F(\zeta^{(0)};0)= \left [
\begin{array}{rrr}
{\partial_{\tau}}^2+1 &  0 & -\cos(\tau) \\
0 & \Omega^2 \D_\tau^2+ B^2  & 0 \\
0 & 0  & -\pi
\end{array}
\right]\label{eq:dF}
\ee
Invertibility of $d_{\zeta}F(\zeta^{(0)};0)$
  can be shown by solving the system of inhomogeneous equations:
\begin{eqnarray}
d_{\zeta} F \cdot \delta\zeta=R
\label{eq51}
\end{eqnarray}
where
\be
\delta\zeta= \left [
\begin{array}{r}
\delta \beta_2 \\
\delta \eta_1 \\
\delta \sigma
\end{array}
\right],\ \ 
R= \left [
\begin{array}{r}
B \\
E \\
\Sigma
\end{array}
\right]
\ {\rm and}\  \Sigma\equiv\int_0^{2 \pi} \cos(\tau) B(\tau) d\tau.\nn\ee
The third equation implies:
\be\delta{\sigma} = -\pi^{-1}\Sigma.\nn\ee
Substitution into the first  
equation yields the equation an $\delta{\beta_2}$:
\be \left(\D_\tau^2+1\right)\delta\beta_2 = B -
 \pi^{-1}\int_0^{2 \pi}\cos(\mu)B(\mu) d\mu\ \cos(\tau),
\nn\ee
which has an even $2\pi$ periodic solution.
Finally the second equation can be rewritten as
\begin{eqnarray}
(\Omega^2 {\partial_{\tau}}^2+ B^2)\delta \eta_1=
E={\sum_{n=0}^{\infty}} \cos(n \tau) e_n
\label{eq52}
\end{eqnarray}
which can then be solved by Fourier series: 
$\delta \eta_1={\sum_{n=0}^{\infty}} g_n\cos(n \tau) $.
We obtain
\begin{eqnarray}
 g_n=(B^2-n^2 \Omega^2)^{-1} e_n.
\label{eq53}
\end{eqnarray}
By {\bf (NR)}, $e_n$ is well-defined for each $n$.
Note also that there is a strictly positive 
 minimum distance of the set $\{n\Omega\}$ to $\sigma(B)$.
Therefore,
\be
\| g_n\|_{l^2(\Z)}\le {\rm dist}\left((n\Omega)^2,\sigma(B^2)\right)^{-1}
 \| E_n\|_{l^2(\Z)},
\nn\ee
from which we have boundedness of the inverse: 
 $\left( \Omega^2 \D_\tau^2+ B^2\right)^{-1}$:
\be
\|\delta\eta_1\|_{H^2(\R;l^2(\Z))}\ \le\ C \| E\|_{L^2(\R;l^2(\Z))}.
\nn\ee

By the implicit function theorem there is a number $\epsilon_0>0$ such that 
 for $\epsilon<\epsilon_0$
 the nonlinear wave equation has a unique periodic solution of the form:
\begin{eqnarray}
y_i(\tau,\epsilon)=K_i+ (\epsilon \cos(\tau) + \epsilon^2 \beta_2(\tau))
  \xi_{\Omega,i}
+ \epsilon \eta_1(i,\tau;\epsilon)
\label{eq54}
\end{eqnarray}
with $\tau=\omega(\epsilon) t$, $\omega^2(\epsilon)=\Omega^2
(1 + \epsilon \sigma(\epsilon))$ and $\beta_2,\eta_1$ $2 \pi$
periodic functions of $\tau$.  Note that $\eta_1={\cal O}(\epsilon)$,
by  the equation $F_2=0$, so that we have 
 (\ref{eq:expansion}). 
This completes the proof of Theorem 5.1.

To prove Corollary 5.1, we note that by hypothesis, the direction of bifurcation
 is spatially odd. It is simple to check that the entire proof goes through
with the spaces ${\cal X}$ and ${\cal Y}$ additionally constrained to consist of
functions $\eta$, such that $\eta(\cdot,-i)= -\eta(\cdot,i)$.
\bigskip

\subsection{Do quasiperiodic solutions bifurcate from the kink?}

If we attempt using the above method of proof to construct quasiperiodic 
solutions, we do not succeed.
 To be specific, suppose we seek to construct a quasiperiodic
 solution which is generated by the
two internal modes $\xi_{\Omega_1}$ and $\xi_{\Omega_1}$. Formally,
 we seek $u_i(t) = u(i,t)$ in the form:
\ba
u(i,t) =  K(i) + \epsilon\sum_{j=1}^2\alpha_j\xi_{\Omega_j}(i)\cos(\Omega_jt) 
 +\epsilon\eta_1(i,t)
\label{eq:formalquasi}\ea 
%
%
Substitution into the nonlinear wave equation gives 
\ba
\left(\D_t^2 + B^2\right)\eta_1 = 
 &&-d^{-2}\epsilon\ 
 {\cal V}\left[ \epsilon\sum_{j=1}^2\alpha_j\xi_{\Omega_j}\cos(\Omega_jt)
 +\epsilon\eta_1\right]\nn\\
&& \ \times \left( \sum_{j=1}^2\alpha_j\xi_{\Omega_j}\cos(\Omega_jt)
 + \eta_1\right)^2.
\nn\ea
Expansion of $\eta_1(i,t)$ as a power series in $\epsilon$,
\be
\eta_1(i,t) = \sum_{j=0}^\infty\epsilon^j\eta_1^{(j)}(i,t),\nn\ee
 leads to a hierarchy of 
inhomogeneous equations of the form:
\be
\left(\D_t^2 + B^2\right)\eta_1^{(j)} = {\cal S}^{(j)}(t)
\label{eq:quasieta1}\ee
where as $j$ increases ${\cal S}^{(j)}(t)$ contains a finite sum of 
terms with time-frequencies
of the form $n_1\Omega_1+n_2\Omega_2$ with $|n_1|\le N_1^{(j)}$, 
  $|n_2|\le N_2^{(j)}$ and $N_1^{(j)},\ N_2^{(j)}$ increasing with $j$.
 Thus it is natural to solve for each $\eta_1^{(j)}$
as a truncated multiple Fourier series:
\be
\eta_1^{(j)} = \sum_{|n_1|\le N_1^{(j)},n_2\le N_2^{(j)}}
 e^{i(n_1\Omega_1+n_2\Omega_2)t}g_{n_1,n_2}^{(j)}.
\nn\ee
Substitution into (\ref{eq:quasieta1}) yields:
\be
\left( B^2 -(n_1\Omega_1 + n_2\Omega_2)^2 \right) g_{n_1,n_2}^{(j)} = 
 {\cal G}_{n_1,n_2}^{(j)},\label{eq:inhom}\ee
where ${\cal G}_{n_1,n_2}^{(j)}$ denotes the term in ${\cal S}^{(j)}(t)$ 
which is proportional to $\exp(i(n_1\Omega_1+n_2\Omega_2)t)$. In order
to ensure the general solvability of (\ref{eq:inhom}) we need that
for any $n_1,n_2\in\Z$, 
\be n_1\Omega_1+n_2\Omega_2\ \notin\ \pm\sigma(B).\nn\ee 
Although it is nongeneric for a frequency in the point spectrum (internal 
mode frequency) to be hit, since the continuous spectrum (phonon band) is an
 interval, generically one has $n_1\Omega_1+n_2\Omega_2\in\sigma_{cont}(B)$,
  for
infinitely many choices of $n_1,n_2$. For such choices the operator
on the right hand side of (\ref{eq:inhom}) 
 is not invertible and these resonant frequencies are
an obstruction to solvability. A closer look at the set of resonances and
their contribution on the dispersive normal form, would give insight into
the {\it lifetime} of such eventually decaying quasiperiodic oscillations. 

\section{\bf Large time behavior in a neighborhood of $K_{gs}$}

In this section we combine the normal form analysis of section 4,
 and the existence theory for time periodic 
 solutions of section 5 
with numerical simulation to get a more detailed picture of the large
time behavior of discrete $\phi^4$ and SG in a neighborhood of the
 ground state kink.  We shall make repeated use of Tables 2 and 4 for $\phi^4$ 
and of Tables 6 and 7 for SG in which the normal form / power equations 
and coherent structures are tabulated. In interpreting these tables
 we recall that $\ve$, introduced in  (\ref{eq3}), measures the size
 of the component of the perturbation about the kink in the internal mode 
subspace.
 In full simulations of the evolution equation $\ve$ is typically of order 
 $10^{-1}$, and therefore some of the decay phenomena anticipated 
 by our analysis, occur on very large time scales 
 ({\it e.g.} $\tau\sim \ve^{-2}, \ve^{-4}$ or longer)
 and are difficult to simulate accurately.  

In each of the various $d-$ ranges we  proceed as follows: 
we mention the relevant coherent structures
(the high energy kink is omitted because it is unstable). These
are all anticipated by the normal form analysis and their existence
is established rigorously in section 5.  We then
 discuss the numerically observed large time behavior for 
 different classes of initial conditions. This gives evidence of the 
attracting nature of the periodic solutions constructed in section 5
 in various regimes  of the discreteness parameter $d$.

As mentioned in the introduction, this is related
to the final stages of the evolution of a propagating kink, pinned to 
a particular lattice site, and its damped oscillation to an asymptotic
state within the Peierls-Nabarro potential. 

\subsection{\bf Discrete $\phi^4$}

\noindent{\bf Regimes II, III and V:}
 The coherent structures of interest are: $K_{gs}$ and $W$.
 Numerical simulations show that the wobbling kink, $W$,
is a local attractor.

\begin{figure}[ptbh]
\centerline{\psfig{width=4in,height=3.9in,file=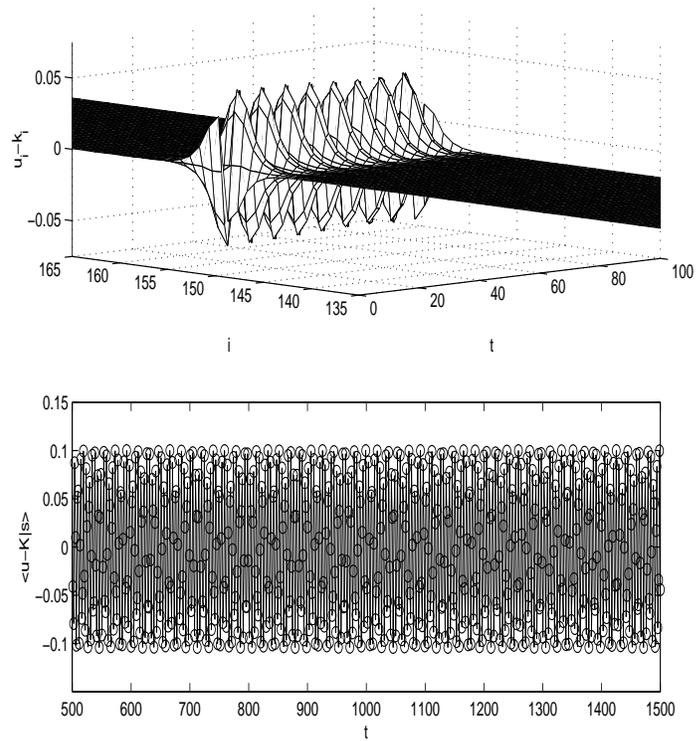}}
\caption{Simulation of discrete $\phi^4$ with $d=0.8$ in Regime IV.
 Upper panel: $u(t)-K_{gs}$ is seen to be approximately
 equal to the (spatially odd) shape mode. Lower panel: Projection of
 $u(t)-K_{gs}$ onto the shape mode, $s$; principal frequency of oscillation 
is approximately $\omega_s$}
\end{figure}

\noindent{\bf Regimes I,IV:}
 The coherent structures of interest are: $K_{gs}$ , $W$ and $gW$.

{\it Experiment 1} (data given by $K_{gs}$ plus a small multiple of the
 Goldstone (even) $g$-mode): solution appears to approach $gW$.

{\it Experiment 2} (data given by $K_{gs}$ plus a small multiple of the 
 shape (odd) $s$-mode): solution appears to approach $W$.

{\it Experiment 3} (data given by a
  general small perturbations of $K_{gs}$):
  $gW$ and $W$ appear to have basins of 
attraction. Although an exact determination of this basin is not
analytically tractable, its projection  onto the internal mode subspace,
 the span of $\{g,s\}$, 
  can be approximated using the internal mode power equations and the 
 explicit information on coefficients from Table 1. 
In particular, we find 
\be {dQ\over dP} = {\omega_g\over\omega_s}\nn\ee
from which one gets the prediction that for data with 
\be |B(0)|^2 < {\omega_g\over\omega_s}|A(0)|^2,\nn\ee
 solutions asymptotically
approach a {\it g-wobbler}, $gW$, while for 
\be |B(0)|^2>{\omega_g\over\omega_s}|A(0)|^2,\nn\ee 
solutions asymptotically
approach a {\it wobbler}, $W$.

\begin{figure}[ptbh]
\centerline{\psfig{width=4in,height=3.9in,file=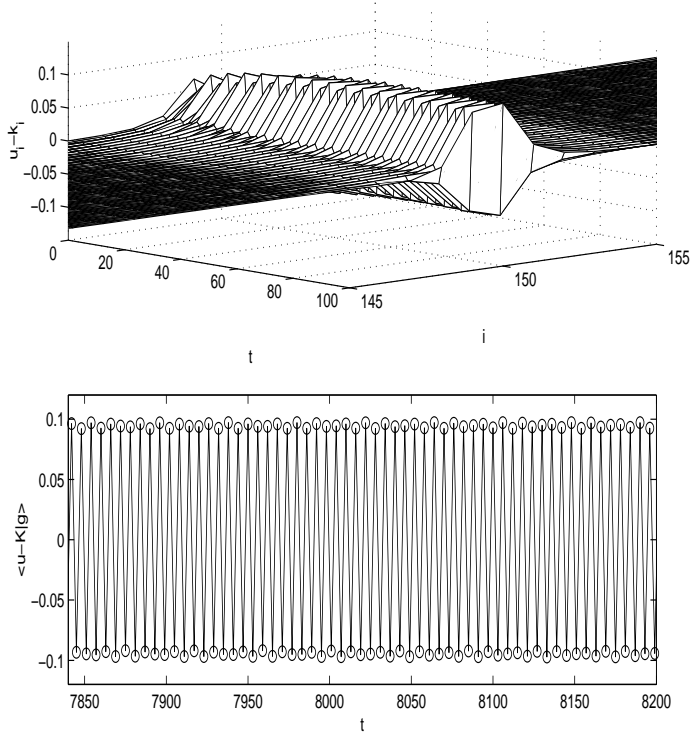}}
\caption{Simulation of discrete $\phi^4$ with $d=0.65$ in Regime IV.
 Upper panel: $u(t)-K_{gs}$ is seen to be approximately 
 equal to the (spatially even) Goldstone mode, $g$. Lower panel: Projection of 
 $u(t)-K_{gs}$ onto the Goldstone mode; principal frequency of oscillation is
 approximately $\omega_g$.} 
\end{figure}

\noindent{\bf Regime VI:}
 The phase portrait of the normal form jumps from dimension $2$ to 
dimension $3$ due to the appearence of a (spatially even) edge mode.
In addition to the ground state kink, $K_{gs}$, there are time-periodic 
 (wobbling solutions) $W$ and $eW$.
 Numerical simulations indicate
the presence of very long lived quasiperiodic oscillations about 
the kink. 
 These appear to be oscillations of the form represented in the first two
terms of  (\ref{eq:formalquasi}). Strong evidence of the eventual decay of 
such oscillations can be seen as follows:
\begin{itemize}
\item In section 5.2 we have shown
that an attempt to construct quasiperiodic solutions breaks down 
due to a high order resonance, {\it i.e.} 
 $n_1\omega_s+n_2\omega_e\in\sigma_{cont}(B)$, where $|n_1|+|n_2|$ is large.
In this particular case, we have $-\omega_s+2\omega_e\in\sigma_{cont}(B)$; 
 see Table 2.
\item Such resonances correspond to obstructions in the power equations to 
equilibrium solutions of the form: $(0,Q,R)$ with both $Q$ and $R$ nonzero. 
The obstructing term is the term $-\ve^4 QR^2$ in the $R$ equation.
\item Linearization of the power equations 
 about any equilibrium point $(0,Q_{eq},0)$ or $(0,0,R_{eq})$ shows that 
each line of equilibria is asymptotically stable, corresponding to the conclusion
 that quasiperiodic oscillations will damp with a resulting asymptotically
periodic state, $W$ or $eW$, depending on initial conditions.  The time scale
of decay of these oscillations, $\tau$, is set by the order in
  $\ve$ of the obstructing 
terms, {\it e.g.} $\tau\sim\ve^{-4}$.
\end{itemize}

\noindent{\bf Regime VII:}
 The coherent structures of interest are: $K_{gs}$ and $eW$. 

\noindent {\it Experiment 1}: For data which is an odd perturbation of $K_{gs}$.
 $K_{gs}$ is the attractor.

\noindent {\it Experiment 2}: For general
data, $eW$ appears to be a local attractor. 

\begin{figure}[ptbh]
\centerline{\psfig{width=4in,height=3.9in,file=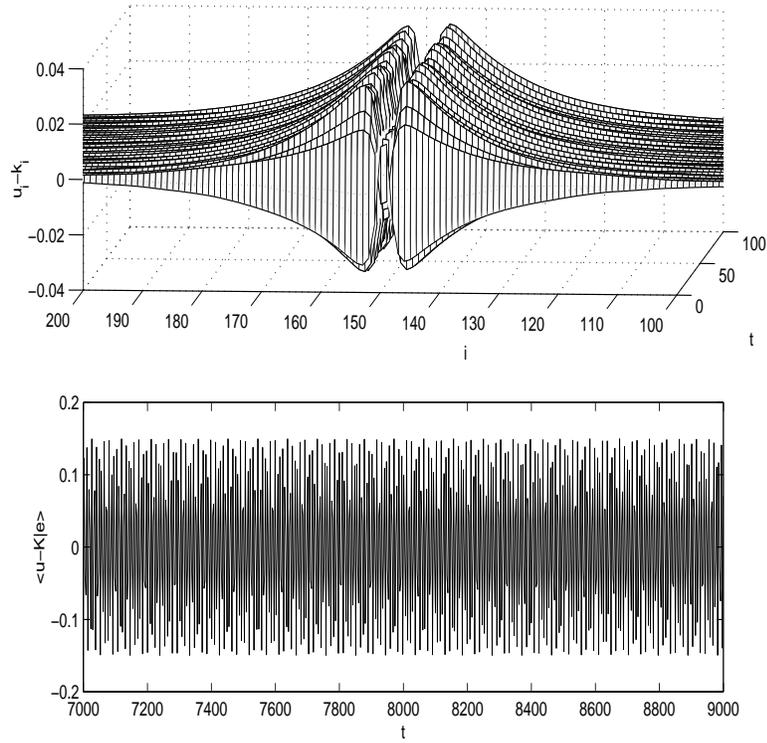}}
\caption{Simulation of discrete $\phi^4$ with $d=1.1$ in Regime IV.
 Upper panel: $u(t)-K_{gs}$ is seen to be approximately
 equal to the (spatially even) edge mode, $e$. Lower panel: Projection of
 $u(t)-K_{gs}$ onto the edge mode; principal frequency of 
 oscillation is approximately $\omega_e$.}
\end{figure}

\noindent{\bf Regime VIII:} The coherent structure of interest is $K_{gs}$. 

\noindent 
 {\it Experiment 1} (data given by $K_{gs}$ plus an small perturbation in the 
direction of the (odd) shape mode): approach to $K_{gs}$ at a rate 
 ${\cal O}(t^{-{1\over2}})$.

\noindent {\it Experiment 2} (data given by $K_{gs}$ plus an small perturbation in the 
direction of the (even) edge mode): approach to $K_{gs}$ at a rate 
 ${\cal O}(t^{-{1\over2}})$.

\noindent
 {\it Experiment 3} (data given by $K_{gs}$ plus an small perturbation in the 
direction of the (even) Goldstone mode): perturbation from $K_{gs}$
 appears to decay but at a different rate and on much longer time scales 
than in Experiments 1 and 2.
\bigskip

\subsection{Discrete sine-Gordon}

Following the model of the previous subsection, we indicate relevant coherent
structures and briefly discuss the dynamics in a neighborhood of the
  ground state kink for the discrete sine-Gordon equation.

\noindent{\bf Regime I:}
 The coherent structures of interest are: $K_{gs}$ and $gW$.
 For general
data, $gW$ appears to be a local attractor.

\noindent{\bf Regime II:}
 The coherent structures of interest are: $K_{gs}$, $W$ and $gW$.
The dynamical behavior is as, for example, in regime IV for the $\phi^4$ model.

\noindent{\bf Regime III:} The coherent structures of interest are $K_{gs}$
and $W$.  Numerical simulations show that $W$ is a local attractor.
 
\noindent{\bf Regime IV:} 
 For all $d\ge d_*$, there are positive integer multiples
  of both $\omega_g$ and $\omega_e$ which 
fall in the phonon band, $\sigma_{cont}(B)$. Therefore, hypothesis {\bf (NR)} of
section 5 is not satisfied for either internal mode, and the  existence theory
of periodic solutions  breaks down. A general perturbation about $K_{gs}$ will excite 
the internal modes and these resonances are responsible for transfer of
energy from the internal modes to $K_{gs}$ and to radiation modes. The asymptotic state
is observed to the ground state kink, with the details of the damped oscillatory approach
to it, being predicted by the various $d-$ dependent normal forms.

Within each subregime of regime IV, the power equations are, up to terms which can be 
shown to be  
negligible for large $t$, of the form 
\ba
P_t &=& -\ve^{2n} P^n\nn\\
Q_t &=& -\ve^2PQ.\nn
\ea
It follows that 
\be
P(t) = P_0 (1+ (n-1)P_0^{n-1}\ve^{2n} t)^{-{1\over n-1}}\nn
\ee
and that $Q(t)$ is decaying at exponentially.
These predicted power laws for $P(t)$ (on a time scale of order $\ve^{-2n}$,
 and the implied very rapid decay of $Q$ 
 are, to a good degree of approximation,
  observed in numerical simulations of the discrete sine-Gordon equation. 
In particular, using the form of the near-identity transformation
($4.24$), we obtain for  the decay rates 
(of initial conditions with general data):
\begin{itemize}

\item $0.86\sim d_*\le d<0.9$:\ $n=3,\ P(t)={\cal O}(t^{-{1\over2}})$ 
$ \rightarrow\ |a(t)|,|b(t)| \sim {\cal O}(t^{-{1\over4}})$   

\item $0.9\le d<0.99$:\  $n=4,\ P(t)={\cal O}(t^{-{1\over3}})$, $\rightarrow 
\ |a(t)|,|b(t)| \sim {\cal O}(t^{-{1\over6}})$ 

\item $0.99\le d< 1.003$:\  $n=5,\ P(t)={\cal O}(t^{-{1\over4}})$, 
 $\rightarrow
 \ |a(t)|,|b(t)|  \sim {\cal O}(t^{-{1\over8}})$ 
\end{itemize}

Note, however, that for purely odd initial perturbations around the
ground state kink in this regime,  the  edge mode projection  
decays as $t^{-1/2}$, as dictated by the $-\ve^2 Q^2$ term.

\section{Summary and open questions}
\setcounter{equation}{0}

\noindent{\bf Summary:}\  In this paper we have considered the  
 large time behavior of solutions to discrete   
  nonlinear wave equations, particularly the discrete sine-Gordon and discrete $\phi^4$
 models, for initial conditions which are a
small perturbation of a stable (ground state) kink.

\noindent 
(1) We have proved in the regime where the discreteness parameter is
sufficiently small, corresponding  
to sufficiently large lattice spacing or weak coupling of neighboring 
oscillators,  
 the existence of various classes of time periodic
solutions. These include the, in the case of the $\phi^4$ the 
 wobbling kinks ($W$), which have the same spatial symmetry
as the kink, and which were anticipated in previous  numerical studies as
well as $e$-wobblers ($eW$) and $g$-wobblers ($gW$), 
 periodic solutions which do not respect 
the spatial symmetry of the kink. In the SG case, it is the $eW$'s that
respect the spatial symmetry while the $gW$'s do not.

\noindent
(2)  We have used the methods of scattering theory and 
( Hamiltonian dispersive) normal forms to 
study the final stages of pinning of a kink to a particular lattice site. 
For large values of the discreteness parameter (``near" the continuum regime),
 the asymptotic state is  a static kink. 
 In contrast, for small values of the 
discreteness parameter, the above time periodic states can be 
 attracting orbits. 
This  is in sharp contrast to the behavior of solutions for the corresponding
 continuum equation in which a (possibly moving) kink is the attractor.
  Our analysis makes clear 
 the relation of broken (Lorentz) invariance to these contrasting dynamics.
Our results give a systematic clarification
 of the physicist's heuristic picture of the dynamics
   of the center of mass of the kink as the effective (radiation-) damped 
 motion of a massive 
 particle in the Peierls-Nabarro potential. 
The approach  to asymptotic state is via a slow damped 
 (periodic or quasiperiodic) 
 oscillation. The damping of the oscillation can indeed be very, very slow,
  and the {\it lifetime} is deducible from the  normal form.

The methods we use are very systematic and 
  general, and apply to many situations \cite{GAFA,SWjsp,SWinvent}
 where the 
dynamical system can be viewed as the interaction of two subsystems: a 
finite dimensional subsystem, here governing the kink and its 
 internal oscillations,
and an infinite dimensional dynamical system,
  here governing dispersive radiation. In particular, one problem of
related interest in which these methods can be readily generalized
is the behavior of pulse-like breathing modes in the discrete non-linear
Schr{\"o}dinger equation $i \psi_{i,t}=-k \delta^2 \psi_i-|\psi_i|^2 \psi_i$. 
Using the monochromatic gauge symmetry of this equation, we can
convert the time periodic problem of the breathing solutions into
a static problem by looking for solutions of the form $\psi_i=
\exp(\sqrt{-1} \omega t) u_i$ and studying their stability in
the frame rotating with the same frequency $\omega$ (the so-called
rotating wave approximation). The excitation and nonlinear 
Lyapunov stability of such states
(see for example footnote 6 in section 2) is established in 
\cite{MIW} This notion of stability, being defined in terms 
of conserved integrals,
is insensitive to the radiative behavior and the asymptotic 
approach to  a ground state. For a recent account of the (numerical)
construction and stability of such modes, see for instance \cite{KRB}.
Then, once the problem has been posed on the rotating wave frame,
it has become a static problem amenable to the same techniques for
the study of dispersive waves and the radiation losses of the 
pulse-like coherent structure as the kink-like structures studies
in this work. Exponentially small effects in the solutions for
this class of systems arising for problems with a perturbed form of 
nonlinearity have  been explored using perturbation theory techniques
\cite{PAK,P2}. These techniques are applicable to a restricted regime of
parameter space because of the adiabatic approximation they entail.
However, to capture, primarily, radiation effects
caused by the discretization in the dynamics of continuum-like 
coherent structures our technique is obviously most appropriate
(and clearly by no means restricted in its applicability)
as illustrated by the analysis given above. What's more, the phenomena
present in such an analysis are in general of the power-law type 
rather than exponential.

\bigskip

\noindent{\bf Open questions:}
There are many related open problems of which we now mention several.

\begin{itemize}

\item Although the normal form analysis anticipates the existence and stability
 properties of 
 various time-periodic solutions, we only have a  rigorous theory concerning their 
existence. Linear stability theory would require an infinite dimensional 
 Floquet analysis, and nonlinear stability theory would require the analogue of
our normal form in a neighborhood of a periodic, rather than static, solution.
Numerically, the asymptotic stability of the time periodic structures
can be concluded from their persistence for long time integrations (to
the extent of our computational power). The periodic structure construction
and stability can also be performed using a limit-cycle type of technique
similar to the one considered in \cite{AUB,MFMAM}. The rigorous stability 
theory however is still a challenging open mathematical problem.

\item A rigorous infinite time theory, as in \cite{SWinvent},
  for discrete nonlinear wave equations with kinks  
  is a challenging  
open question. The problem is challenging even in the continuum case, 
 where the kink is the sole 
 attractor. In this case, the normal form yields the correct asymptotic rate of decay
 \cite{Manton,PSW}. However, due to the strength of the interaction (in space dimension
one) of the dispersive and bound state components,
  the deviation of the full solution, $u(t)$, from 
 the kink plus its internal mode modulations is {\bf not} asymptotically free,
 {\it i.e.} not a free
  dispersive solution of the linear Klein-Gordon equation.
 Rather, it requires a solution-dependent
  logarithmic in time corrected phase \cite{PSW},  characteristic
 of long range scattering problems.

\item Of interest would be more detailed long time simulations which would 
further elucidate the large time dynamics and assist in mapping out basins of
attraction for various periodic states.  

\item Our study gives a detailed picture of the final stages of the pinning of 
a kink  to a particular lattice site. However, the early stages of a kink plus 
{\it large} perturbation propagating in a lattice have some of the same features 
of the regime we consider. 
 Namely \cite{PK}, as the kink's center of mass moves under the influence of the 
Peierls-Nabarro potential, it executes repeated acceleration and slowing with 
 an approximate PN frequency. This oscillation appears to resonate with 
continuous radiation modes, resulting in an  emission of energy,
 deceleration of the kink and eventual capture of the kink 
by a particular lattice site. It would be of interest to extend and 
 apply the ideas of the current
paper to this problem.

\item In this paper we have analyzed the problem of radiative
effects of discreteness for the $2-\pi$ kinks within the P-N barrier
(for a recent review on the effects of discreteness as well as the physical 
applications of a number of models similar to the ones considered here
see \cite{BRKI}). It is well-known however that in the SG lattices
multiple kinks ($2 n \pi$ kinks in general) can also exist. For these
kinks first mentioned in \cite{PK} the static stability picture was
analyzed in \cite{BCK}. There are a number of open problems regarding the
behavior of such modes:
\begin{enumerate}
\item Their motion prior to pinning is only very slightly effected
by radiative phenomena \cite{PK}. This phenomenon hasn't been accounted for
satisfactorily to the best of our knowledge to date. In fact, this forms
a part of a more general problem concerning the motion of coherent structures
in lattices. It is known that under certain conditions and for potential
different than the ones considered here \cite{CMPVV,IK} there exist
lattice systems with travelling wave solutions. Hence, it would be
very interesting to elucidate the general conditions under which 
such lattice systems support travelling wave solutions (or equivalently
under which conditions solutions to the advance-delay equations of the 
travelling wave frame exist). 
\item When trapped (eventually) in the PN barrier these moving $4-\pi$ 
kinks also radiate in a way that can be captured by the analysis of
this paper. The problem there will have more (as shown in \cite{BCK})
internal modes bifurcating from the continuum and hence will be more
complicated but it can still be analyzed
in the way presented above. 
\end{enumerate}
\end{itemize}


{\Large\bf Acknowledgements}

Thanks are expressed to N.J. Balmforth, A.R. Bishop, 
J.L. Lebowitz, C.K.R.T. Jones, Y.G. Kevrekidis, H. Segur
  and A. Soffer for stimulating
discussions and for suggesting some important references.
P.G.K. gratefully acknowledges assistanship support from the 
Computational Chemodynamics Laboratory of Rutgers University,
fellowship support from the ``Alexander S. Onasis'' Public Benefit
Foundation and also partial support from DIMACS.

\end{document}